\documentclass[
preprint,
aps,prd,
nofootinbib,
superscriptaddress,
showpacs,
tightenlines,
amsmath,
amssymb
]{revtex4}
\usepackage{bm}
\usepackage[dvips]{color,graphicx}

\begin{document}
                               

\title{Nuclear Medium Modifications of Hadrons from Generalized Parton 
Distributions.}

\author{S.~Liuti}
\email[]{sl4y@virginia.edu}

\author{S.~K.~Taneja}
\email[]{skt6c@virginia.edu}

\affiliation{University of Virginia, Charlottesville, Virginia 22901, USA.}


\begin{abstract}
We study the structure of generalized parton distributions 
in spin $0$ nuclei within a microscopic approach for nuclear dynamics.
GPDs can be used on one side as tools 
to unravel the deep inelastic transverse structure of nuclei  in terms of 
both transverse spatial and transverse momentum degrees of freedom. On the
other, one can obtain information on GPDs themselves by observing how they 
become modified in the nuclear environment.
We derive the structure of the nuclear deeply virtual Compton scattering 
tensor and generalized parton distributions 
at leading order in $Q$ in a field-theoretical framework. 
The nuclear generalized parton distributions are calculated using a 
two step process -- the convolution approach -- where the scattering 
process happens from a quark inside a nucleon, itself inside a nucleus,
disregarding final state interactions with both the 
nuclear and nucleon debris.
We point out that details of the nuclear long range interactions such as two-body currents,
can be disregarded compared to the deep inelastic 
induced modifications of the bound GPDs. 
We show how the pattern of nuclear modifications predicted, and in 
particular the deviations of off-shell effects from the longitudinal convolution
provide clear signals to be sought in experimental measurements. 
Finally, we find interesting relationships by studying Mellin moments in nuclei:
in particular we predict the $A$-dependence for the $D$-term of GPDs within a microscopic approach, 
and the behavior
with $t$ of the total momentum carried by quarks in a nucleus. The latter 
provides an important element for the evaluation of nuclear hadronization 
phenomena which are vital for interpreting current and future data 
at RHIC, HERMES and Jefferson Lab.
\end{abstract}

\maketitle
\section{Introduction}
\label{intro_sec}
Generalized Parton Distributions (GPDs)
have sensibly transformed our views and approaches on probing hadronic
structure, since they were first introduced \cite{DMul1,Ji1,Rad1}. 
They provide, in fact, a framework 
to describe in a partonic language 
the orbital angular momentum carried by the nucleon's 
constituents \cite{Ji1}. In addition, they give invaluable information
on the partonic distributions in the transverse direction with 
respect to the large longitudinal momentum in the reaction \cite{Bur}.

Recent developments focused on the relationship between
GPDs, the Impact Parameter dependent Parton Distributions 
(IPPDs), and transversity \cite{Bur,Diehl03}.
Several usually unaccounted for spatial observables including the transverse
{\it location} of partons carrying a longitudinal fraction of momentum $x$,  
the interparton separation, and the radius of
the proton, have been expressed in terms of both IPPDs and GPDs.
The intrinsic transverse momentum distributions (or the Unintegrated Parton 
Distributions, UPDs), and their relation to 
the IPPDs, and GPDs, as well as to the nucleon Form Factors (FFs), 
and to the inclusive structure functions, can also be studied within the same
context \cite{Diehl03,LiuTan1,DieKro,ChaMuk}.
A new perspective on GPDs as ``tools'' to study 
transversity in hadronic systems has just now begun 
to unravel \cite{BelJiYuan}.  

The most straightforward method to access GPDs experimentally is to use 
a class of 
exclusive $ep$ hard scattering experiments that proceed through the 
exchange of a highly virtual photon, the final state being the proton, or
a proton resonance, and a real photon, or a vector meson.     
Initial experimental results exist to date for  
Deeply Virtual Compton Scattering (DVCS) off a proton target, a process that has also 
become the prototype for describing different aspects of GPDs (see {\it e.g.} 
Ref.\cite{Die_rev} for a review). Nevertheless, it is also clear that 
by using only this procedure, the mapping of 
the GPDs in a significant fraction of the phase space, and for a range 
of values
of the process' scale might be quite lengthy.

GPDs were recently also measured through DVCS off nuclear targets 
\cite{Hasch}. The study of nuclear targets is particularly important as they 
provide a laboratory where additional information
can be obtained on these elusive observables. Exploratory studies and estimates 
concentrating on the passage of nucleons through the nuclear medium, 
possibly generating Color Transparency, were performed in \cite{PirRal_DVCS,BurMil,LiuTan1}. 
Another equivalently interesting possibility is 
that by keeping track of possible 
transverse deformations of nucleons in the nuclear medium, 
one can test the behavior of the GPDs themselves.
In this paper we present a first 
investigation of this idea using a microscopic description of the nucleus.

The most clearcut experimental evidence that nucleons have a different 
deep inelastic structure
inside the nuclear medium than when isolated, is represented by the EMC 
effect \cite{EMC}.
The EMC effect is the observation of a different behavior 
of the inclusive electron/muon nuclear cross section per nucleon 
with respect to 
the free nucleon one (see \cite{Tho} for a review). 
These differences  
appear at intermediate values of Bjorken $x$ ($x_{Bj} = Q^2/2M_N\nu$,
$Q^2$ being the four-momentum transfer, $M_N$ the nucleon mass, $\nu$ the
energy transfer), 
namely, $0.1 \lesssim  x_{Bj} \lesssim 0.7$, where they cannot be 
attributed either to expected coherent effects such as nuclear shadowing
($x_{Bj} << 0.1$), or to Fermi motion ($x_{Bj} \rightarrow 1$).    
Throughout the years since the first discovery of the EMC effect
an increasingly coherent picture has emerged confirming that
even in a Deep Inelastic Scattering (DIS) process 
characterized by a high locality of the probe-target interaction
region, 
nucleons cannot be treated as free. 
Their interactions 
are instead important, they can bear as important consequences as 
being responsible for modifications of {\it e.g.} the chiral condensate,  
the color string tension, or the confinement scale  
in cold nuclear matter \cite{Pirner},  
\footnote{Interactions become more and more important at finite
temperature where non-perturbative properties of QCD, {\it i.e.} its vacuum, 
and confinement, are expected to be modified  
(see {\it e.g.} \cite{Wam}).}
and, therefore, play a key role in understanding the 
origin of hadronic mass. 

Despite a quite general consensus on this picture,  
the form these interactions can take is still very model dependent, 
ranging from 
new applications of effective theories \cite{Mil_A,Tho_A} to 
extensive studies of the impact of modifications of the confinement 
scale, or of the string tension for bound nucleons 
(\cite{Pirner} and references therein).  
The extension to finite temperature
measurements is possible but challenging, 
however at the light of current and future relativistic heavy ion collisions, 
and $eA$ collisions  it is 
important to better define the connection with nuclear medium modifications 
that can be considered as precursors to the finite $T$ ones. 
The avenue that we espouse in this paper, is in this spirit.  
In particular, we make use of GPDs as a tool
to access transverse coordinates and momenta in the nucleus. 
Transverse degrees
of freedom, in turn, play a special role in quark re-interactions
\cite{BroHwaSch} and we expect them to play a significant role
in nuclear medium modifications, and in hadronization \cite{Pirner}. 
In our description of inclusive DIS from a nucleus, 
re-interactions are parametrized as off-shell effects. 
{\em Kinematical} off-shell effects in a nucleus derive from the 
modification of the relation between
the struck quark's transverse momentum, $k_\perp$, and its virtuality, $k^2$, 
produced by the $P^2$ dependence 
of the bound nucleon's structure function. 
It is well known that the account of kinematical off-shell effects 
alone is an oversimplified picture that produces a violation of Adler's
sum rule \cite{EFP,KPW}, and that {\em dynamical} nucleon off-shell effects
should also be introduced. 
An important aspect of the approach proposed here is that, 
by restoring Adler's sum rule in a nucleus,
it provides a description of the EMC effect that, at variance 
with the naive (on-mass-shell) binding models, can simultaneously, and
quantitatively, 
reproduce both the $x_{Bj}$ and $A$ dependences of the data. 

In Section \ref{def_sec} we define our general framework; 
in Section \ref{models} we calculate GPDs in nuclei with offshell 
parametrizations. 
We subsequently evaluate in Section \ref{moments-sec} 
the role off-shellness in the second Mellin moments of the nuclear GPDs.
Finally, in Section \ref{conc_sec} we give our conclusions and outlook on 
future developments. 

\section{General Framework}
\label{def_sec}

In this Section we present the formalism used to derive GPDs in 
nuclei in terms of their constituents, the protons and neutrons, 
including the effects of nucleon binding. Nuclear GPDs were at first calculated
for the deuteron \cite{CanPir}. In Ref.\cite{KirMul} the general
structure of the nuclear hadronic tensor within OPE was carefully evaluated for
spin 0, 1/2, and 1 nuclei, 
however, a rather crude estimate of dynamical effects was given. 
The latter were evaluated in  
\cite{FreStr,StrWei} only in the low $x$ region, dominated by nuclear 
shadowing. Quantitative calculations for the valence region were performed in
\cite{Sco} for $^3$He, within the (longitudinal) convolution model, 
{\it i.e.} disregarding off-shell effects, and transverse degrees of freedom.  

The results we present here are also valid in the valence region 
($x \geq 0.2$). Our evaluation of nuclear effects 
is extended to larger nuclei than 
the deuteron and $^3$He, namely to complex nuclei such as $^4$He, $^{12}$C, 
etc..., that are close in atomic number to Ne and Xe, 
used in HERMES experiments \cite{Hasch}. 
Furthermore, we include nucleon off-shell effects.
These play an essential role in the determination of nucleon
deformations in the nuclear medium.     

\begin{figure}
\includegraphics[width=7.5cm]{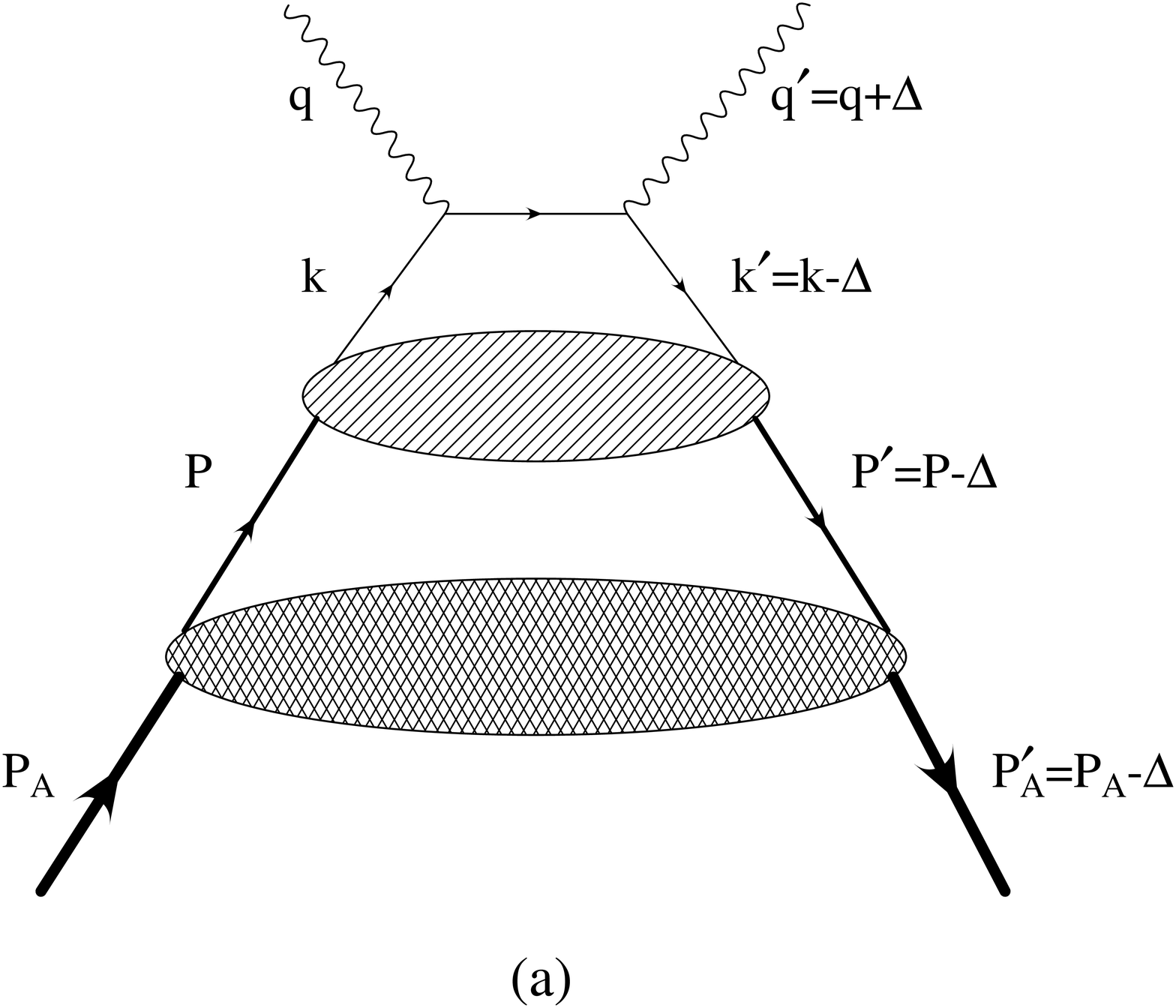}
\includegraphics[width=6.cm]{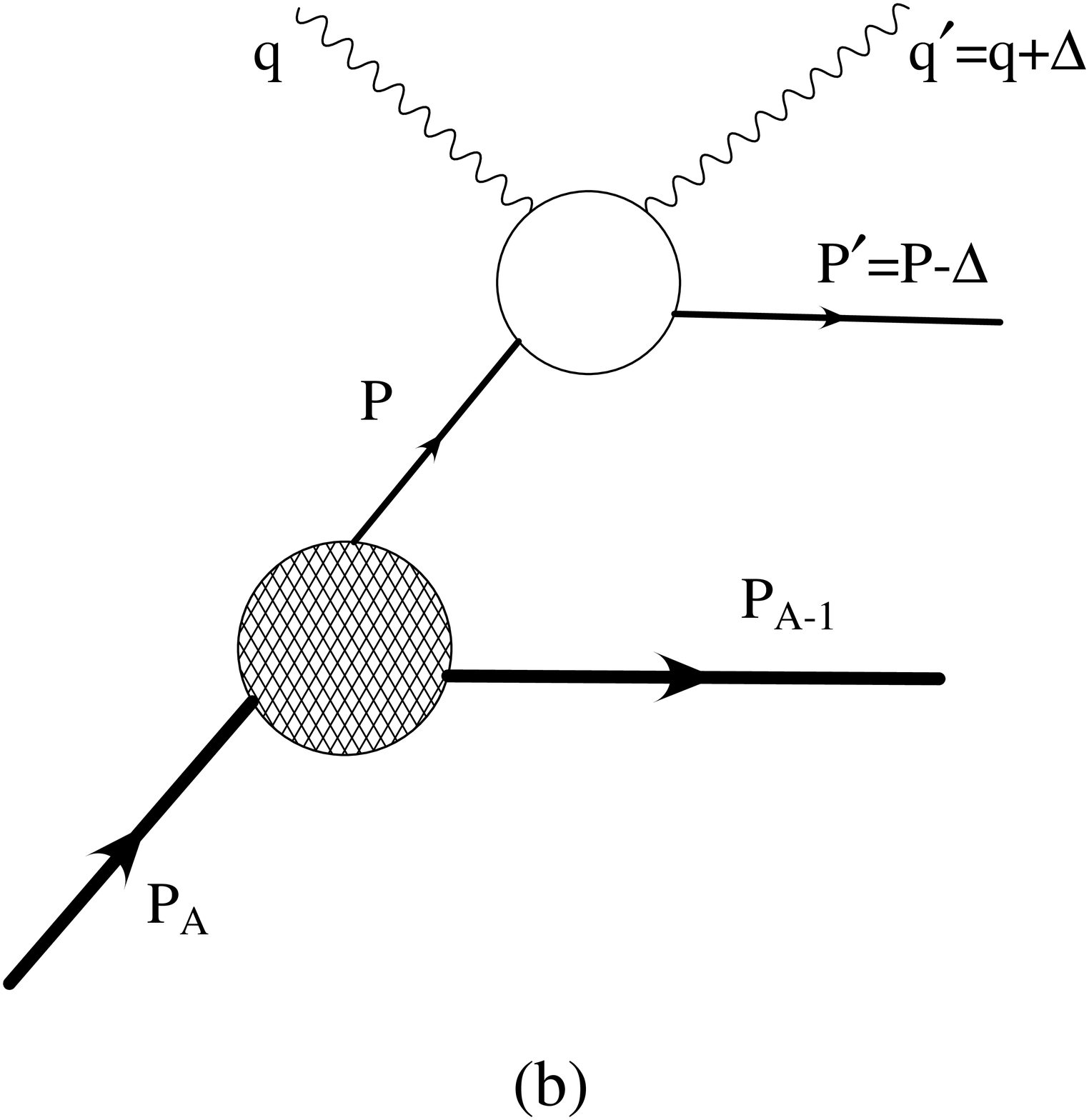}
\caption{Amplitude for DVCS from a nuclear target 
at leading order in $Q^2$. {\bf (a)} Coeherent process; {\bf (b)} Incoherent
process. For the incoherent process the nucleon handbag diagram is also assumed.} 
\label{fig1}
\end{figure}

We consider DVCS off nuclei  
at leading order in $Q^2$ (twist-2). 
The diagrams in Fig.\ref{fig1} 
represent the two possible mechanisms through which DVCS off nuclei occurs: 
{\bf (a)} coherent scattering, 
where the
virtual photon with four-momentum $q$ scatters from a nuclear target
producing a real photon $q^\prime$, the final nucleus $A$ 
recoiling as a whole; {\bf (b)} incoherent scattering, where 
the final nucleus breaks up into a nucleon and an $A-1$ system,
with final invariant mass squared, 
$M_A^{* 2} \approx M_A^2 +$ ``soft excitation terms''. 
A similar treatment applies to 
the other hard exclusive reaction used as a probe of GPDs
in nuclei, namely hard meson production.
In what follows we calculate the ``double layer'' diagram 
(Fig.\ref{fig1}a) because of its relation to the forward
EMC effect. 
Current experimental data \cite{Hasch} contain in principle both 
contributions, a separation of the two being however feasible in 
the near future. In the present paper we are concerned with the 
coherent contribution only, because of its relation to 
the (forward) EMC effect. 
A comparison of the contributions of coherent and incoherent scattering
will be carried out in \cite{LiuTan3}. 
Other components besides the nucleon 
($\Delta$'s, $6$-quark bags, etc...) 
could in principle be similarly described by the lower part of 
the diagram. 
In this paper we take the point of view that
both the $x_{Bj}$ and $A$ dependences of 
the EMC effect can be explained quantitatively 
within a ``minimal'' approach involving only nucleons. 
The nucleon's off-shellness however has to be considered explicitly, as
it plays the role
of a parametrization of soft interactions among 
nucleons (or quarks and nucleons debris) during the scattering process.

We denote by $P_A$, $P$, $k$, the nuclear, the active nucleon's and  
active quark's four-momenta, respectively.  
$P_A^\prime=P_A-\Delta$, $P^\prime=P-\Delta$, and 
$k^\prime=k-\Delta$ are the final nuclear, nucleon's and quark's momenta,
respectively; $q$ is the virtual photon momentum, and $q^\prime = q+\Delta$,
the outgoing photon momentum;  $\Delta^2 \equiv t$. Notice that 
$\Delta = k - k^\prime  \equiv P - P^\prime \equiv P_A - P_A^\prime$, 
in the absence of final state interactions. 
 
Two distinct formulations of GPDs are given in the literature, that
differ in the ``choice of the defining four-vector'' \cite{GolMar}.
Since both formulations turn out to be useful in a complementary way,
in the following we list the kinematical variables and definitions
for both. 
 
In set {\bf (1)} the kinematical variables are defined 
with respect to
the average of the target's incoming and outgoing momenta following 
\cite{Ji1}. These are: 
$\overline{P}_A= (P_A+P_A^\prime)/2$, and $\overline{P}= (P+P^\prime)/2$,
for the nucleus and for the off-shell ($P^2 \neq M^2$) nucleon, 
respectively. 
One has :
\footnote{We use the notation: 
$ \displaystyle p^\pm = \frac{1}{\sqrt{2}}(p_o \pm p_3)$, with
$\displaystyle (pk) = p^+k^- + p^-k^+  - p_\perp \cdot k_\perp$. }    
\begin{subequations}
\begin{eqnarray}
x & = & (k^+ + k^{\prime +})/2(\overline{P}_A^+/A)   \\ 
y & = & \overline{P}^+/(\overline{P}_A^+/A)  \\ 
x_N & = & x/y \equiv (k^+ + k^{\prime +})/2 \overline{P}^+   \\  
\xi & = &  \Delta^+/2(\overline{P}_A^+/A) \\
\xi_N & = & \Delta^+/2\overline{P}^+ \equiv \xi/y 
\end{eqnarray}
\label{kin1}
\end{subequations}
The resulting distributions are real, and they are even functions 
of the variables $\xi$, $\xi_N$.

If the variables are defined with respect to 
the incoming nucleon momentum as in \cite{Rad1} one has:  
\begin{subequations}
\begin{eqnarray}
X & = & k^+/(P_A^+/A) \\ 
Y & = & P^+/(P_A^+/A) \\
X_N & = & X/Y \equiv  k^+/P^+ \\ 
\zeta & = &  \Delta^+/(P_A^+/A) \\
\zeta_N & = & \Delta^+/P^+ \equiv \zeta/Y
\end{eqnarray}
\label{kin2}
\end{subequations}
The structure functions defined in 
terms of these variables are more straightforwardly connected 
with the DIS parton distributions. 

GPDs written in terms of the two sets of variables 
can be translated into one another. 
The graph in Fig.\ref{fig1}, in fact, simultaneously defines 
three configurations where:
{\it i)} both $k$ and $k^\prime$ are quarks; 
{\it ii)} both $k$ and $k^\prime$ are antiquarks;
{\it iii)} $k$ is a quark (antiquark) and $k^\prime$ is an antiquark 
(quark).   

In the quark region, 
that is the main interest for this paper because it is where 
the EMC effect in DIS occurs, 
$\xi < x <1 $, $\zeta < X <1$, and the following 
relations hold:
\begin{subequations}
\begin{eqnarray}
X & = & \frac{x+A\xi}{1+\xi} \\
Y & = & \frac{y+A\xi}{1+\xi} \\
X_N & = & \frac{x+\xi}{y+\xi} \\
\zeta & = & \frac{2\xi}{1+\xi/A} \\
\zeta_N & = & \frac{2\xi_N}{1+\xi_N} 
\end{eqnarray}
\label{kin3}
\end{subequations}
Similar relations can be derived in the antiquark sector ($-1<x<-\xi$ and $\zeta < X <1$). 
\footnote{The region $-\xi < x < \xi$ requires a more detailed description and it
is beyond the scope of this paper. See however \cite{GolMar} for the nucleon case.}

\subsection{Nucleon}
We first summarize the main results in the free nucleon case.
The amplitude for Deeply Virtual Compton Scattering (DVCS) is defined as: 
\begin{eqnarray}
T^{\mu \nu}(P,\Delta,q)= i \int d^4 y \, e^{iq.y}
\langle P^\prime |T \left( J^{\mu}(y)J^{\nu}(0) \right) |P \rangle.
\end{eqnarray}
Notice that we will only assume, and not write explicitly, 
the dependence of the various terms 
on the virtual photon's momentum, $q$.
The factorization theorem for hard scattering processes (Fig.\ref{fig1})
allows one to separate the hard part, calculated using QCD Feynmann rules, 
from the soft hadronic matrix element, ${\cal M}$:
\footnote{We do not include for the time being either radiative corrections, 
or \protect${\cal O}(1/Q^2)$ terms.}
\begin{equation}
T^{\mu \nu}  =  -i \int \frac{d^4 k}{(2\pi)^4} 
{\rm Tr} \left[ \left(
\frac{\gamma^\nu i(\not\!k + \not\!q)\gamma^\mu}{(k+q)^2+i\epsilon} +
\frac{\gamma^\mu i(\not\!k +\not\!\Delta - \not\!q)\gamma^\nu}{(k+\Delta-q)^2+i\epsilon} \right) \, {\cal M}(k,P,\Delta) \right].
\label{h_tensor}
\end{equation}
${\cal M}(k,P,\Delta)$ is an off-forward correlation function:
\begin{eqnarray}
{\cal M}_{ij}(k,P,\Delta)=\int d^4 y \, {e^{ik.y}}
\langle P^\prime | \overline{\psi}_{j}(-y/2)\psi_{i}(y/2) |P \rangle,
\label{matrix}
\end{eqnarray}
with Dirac indices $i,j$ written out explicitly. Notice that we have written
the argument of $\psi$ choosing the ``symmetrical'' 
formalism of \cite{Ji1}.   
By taking the Bjorken limit, and by projecting out the
dominant contribution 
corresponding to the transverse virtual photon polarization, one
obtains:
\begin{equation}
T^{\mu\nu} = \left(- g^{\mu\nu} + 
\frac{\widetilde{P}^\mu \widetilde{P}^\nu}{\widetilde{P}^2}
 - \frac{q^\mu q^\nu}{Q^2} \right) {\cal F}_T 
\label{tensor_trans}
\end{equation}
where $\displaystyle 
\widetilde{P}_\mu=\overline{P}_\mu + (\overline{P}\cdot q/Q^2) q_\mu $.
The off-forward structure function can then be extracted
from ${\cal F}_T$ as:
\footnote{Notice that we adopt the
axial gauge, although results can be cast in a form 
highlighting gauge invariance \cite{Ji1}.}  
\begin{eqnarray}
{\cal F}_T(P,\Delta)  =   
\int_{-1}^1 dx \left(\frac{1}{x-\xi/2 + i \epsilon } - 
\frac{1}{x + \xi/2 - i \epsilon } \right) F(x,\xi,t) ,
\end{eqnarray}
with:
\begin{eqnarray} 
F(x,\xi,t) =
\frac{1}{2 \overline{P}^+} \left[ {\overline{U}(P',S')}\left( \gamma^+ H(x,\xi,t)+
\frac{i \sigma^{+ \mu} \Delta_\mu}{2M} E(x,\xi,t) \right)U(P,S) \right]
\label{sfn}
\end{eqnarray}
Eq.(\ref{sfn}) defines the GPDs, $H$ and $E$, for the unpolarized scattering case \cite{Ji1}. 
$H$ and $E$, and the generalized distributions, 
${\cal F}_\zeta(X,t)$ and ${\cal K}_\zeta(X,t)$  
introduced similarly by Radyushkin using  
Double Distributions (DD) in \cite{Rad1} describe the same physics, however 
the connection between the two different sets of kinematical variables, 
and physical regimes spanned needs to be specified. 

In this paper, 
we adopt the convention first illustrated in \cite{GolMar}, that is we define 
$H_{q(\overline{q})}(X,\zeta,t)$, and 
$E_{q(\overline{q})}(X,\zeta,t)$,
\footnote{For completeness, the correspondence with the notaion in \cite{GolMar} is:
$H_{q(\overline{q})} \equiv {\cal F}_{q(\overline{q})}(X,\zeta,t)$, and 
$E_{q(\overline{q})} \equiv {\cal K}_{q(\overline{q})}(X,\zeta,t)$ } 
directly in terms of $H$ and $E$, as from Eq.(\ref{sfn}), 
but: {\it i)} Changing the kinematical variables to 
set (\ref{kin2}), with $A=1$ for the free nucleon; {\it ii)} Defining explicitly the contributions of
quarks and antiquarks. Because we are interested in the region dominated 
by DGLAP evolution (and $\xi>0$), we consider the following mapping of both regions: $\xi<x<1$ describing scattering 
from a quark, and $-1<x < -\xi$ describing scattering from an antiquark onto $\zeta <X<1$: 
\begin{subequations}
\label{Fq}
\begin{eqnarray}
(1-\zeta/2)H_q(X,\zeta,t) & = &   H(x,\xi,t)  \; \; \; x> \xi, \;  \; X>\zeta
\\
(1-\zeta/2)H_{\overline{q}}(X,\zeta,t) & = & -H(x,\xi,t)  \; \; \;   x <-\xi, \;  \; X>\zeta
\end{eqnarray}
\end{subequations}
and 
\begin{subequations}
\begin{eqnarray}
(1-\zeta/2)E_q(X,\zeta,t) & = &  E(x,\xi,t) \; \; \; x> \xi, \;   \; X>\zeta
\\
(1-\zeta/2)E_{\overline{q}}(X,\zeta,t) & = & -E(x,\xi,t) \; \; \; 
x <-\xi, \;   \; X>\zeta,
\end{eqnarray}
\end{subequations}
where the factor $1-\zeta/2 \equiv \overline{P}^+/P^+$ is due to
the difference in ``reference momenta'' (see Eqs.(\ref{kin1}). 
The new definitions are best suited both for 
describing the convolution with nuclear variables (see below), as well as 
for perturbative evolution.    
Notice that for $\xi,\zeta =0$, for instance, one maps 
the entire $X \in [0,1]$ domain onto the $x \in [-1,1]$ domain.   
As a result, $H_q(X,\zeta,t)$ and $H_{\bar{q}}(X,\zeta,t)$, 
can be expressed in terms of the Non-Singlet (NS), or Valence (V), 
and Singlet (S) contributions as: 
\begin{subequations}
\begin{eqnarray}
H^V& = & \sum_{q=u,d} H_q(X,\zeta,t) - H_{\overline{q}}(X,\zeta,t)
\\
H^S & = & \sum_{q=u,d,s, ...} H_q(X,\zeta,t) + H_{\overline{q}}(X,\zeta,t)
\end{eqnarray} 
\end{subequations}
$H_V$ and $H_S$ become the usual 
valence and singlet quark distributions in the forward limit.

The proton and neutron GPDs are obtained from $H_{q(\overline{q})}$ as:
\begin{subequations}
\begin{eqnarray}
H^p & = & \frac{2}{3}H_u - \frac{1}{3} H_d - \frac{1}{3} H_s \\ 
H^n & = & - \frac{1}{3}H_u + \frac{2}{3} H_d - \frac{1}{3} H_s.
\label{Hpn}
\end{eqnarray}
\end{subequations}
Similar formulae hold for $E_{q({\overline q})}$.
In what follows we assume that isospin invariance holds in a nucleus
as well.  

In Section \ref{models} we calculate GPDs using an effective theory
for the proton, with a quark and a diquark component, as the lowest
light-cone Fock state wavefunction.  


\subsection{Nucleus}
 
We consider now the case in which the nucleon is bound in a nucleus
with mass number $A$ and mass $M_A$. The off-forward nuclear amplitude
can be calculated analogously to Eq.(\ref{h_tensor}) 
by convoluting the truncated 
off-forward nucleon amplitude, $T_{\mu\nu}^N$, with the nuclear matrix 
element, ${\cal M}^A(P,P_A,\Delta)$:   
\begin{equation}
T_{\mu\nu}^A(P_A,\Delta) = \int \frac{d^4 P}{(2\pi)^4} 
T_{\mu\nu}^N(k,P,\Delta) \, {\cal M}^A(P,P_{A},\Delta) ,
\label{A_tensor}
\end{equation}
where: 
\begin{equation}
{\cal M}_{ij}^A(P,P_{A},\Delta) =\int d^4 y \, e^{iP \cdot y}
\langle P_A^\prime | \overline{\Psi}_{A, j}(-y/2)\Psi_{A, i}(y/2) |P_A \rangle.
\end{equation} 
$T^N_{\mu\nu}(k,P,\Delta)$ is the hadronic tensor for a bound nucleon
(we consider from now on the isoscalar combination, $N=(p+n)/2$).

The structure function can be extracted from 
Eq.(\ref{A_tensor}) similarly to the procedure used to 
obtain Eq.(\ref{sfn}) from Eq.(\ref{h_tensor}), leading to:
\begin{equation}
F^A(X,\zeta,t) = \int \frac{d^4 P}{(2\pi)^4}
F^N_{OFF}(X_N,\zeta_N,P^2,t) \, {\cal M}^A({P},{P_{A}},\Delta), 
\label{convo1}
\end{equation}
where all kinematical variables have been defined in 
Eqs.(\ref{kin1},\ref{kin2},\ref{kin3}). 
Notice that the bound nucleon is
off its mass shell, namely: 
$P^2 \equiv P^2(X,P_T^2) \neq M^2$, and consequently 
$F^{N}_{OFF}(X_N,\zeta_N,P^2,t) \neq F^N(X_N,\zeta_N,t)$.
$F^N(X_N,\zeta_N,t) = (F^p + F^n)/2$ is the nucleon, or isoscalar, combination of 
GPDs, obtained using Eqs.(\ref{Hpn}).

The form of ${\cal M}_A$ depends on the spin of the nucleus. 
In the next Section we evaluate Eq.(\ref{convo1}) for a nucleus 
with spin 0.

\section{Evaluation of Nuclear GPDs}
\label{models}
\begin{figure}
\includegraphics[width=11.cm]{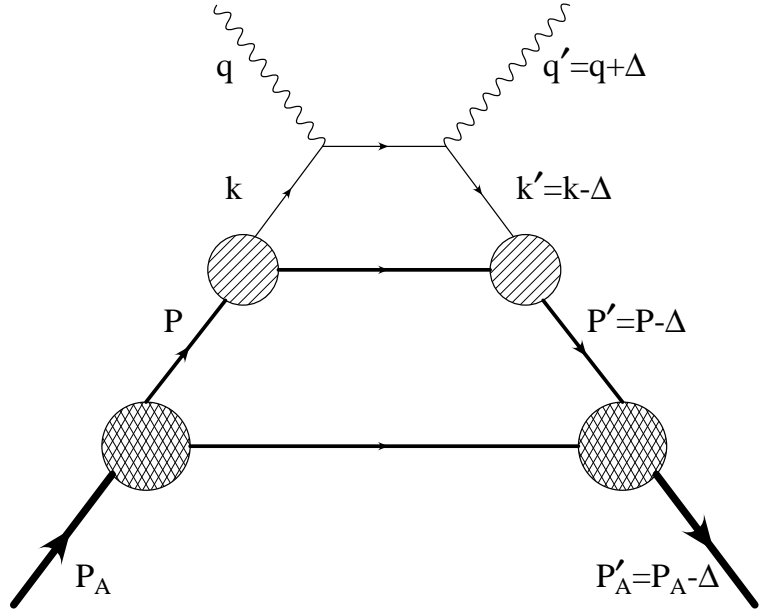}
\caption{Off-forward nuclear double convolution diagram. The nucleon is described
by a quark-diquark model, and the nucleus is treated within the Impulse Approximation.} 
\label{fig2}
\end{figure}

The evaluation of Eq.(\ref{convo1}) requires
a mechanism for describing nucleon off-shellness.
We adopt an effective theory described in Fig.\ref{fig2},  
where the coupling at the nucleon vertex is: 
$g(k^2)\overline{\psi}_N \psi_q \phi$,
$\psi_N$, $\psi_q$ and $\phi$ being the wave functions of the nucleon, $P$, the 
active quark, $k$, and the spectator diquark, $k_X$, respectively.
Within this model, that can be considered a realization of the lowest
Fock component for the light cone wave function, 
the active quark's virtuality, $k^2_\mu \neq m_q^2$, is related 
to the intrinsic transverse momentum, and either variable can be integrated
over to obtain the forward parton distribution (see {\it e.g.} 
\cite{JakMul} and references therein).     
Similarly, one can treat the ``active nucleon''
within a relativistic nuclear effective theory
(lower part of diagram in Fig.\ref{fig2}). 
Different versions of the spectator model were used  
to calculate the (forward) EMC effect \cite{GroLiu,KPW,AKL}.

\begin{figure}
\includegraphics[width=8.cm]{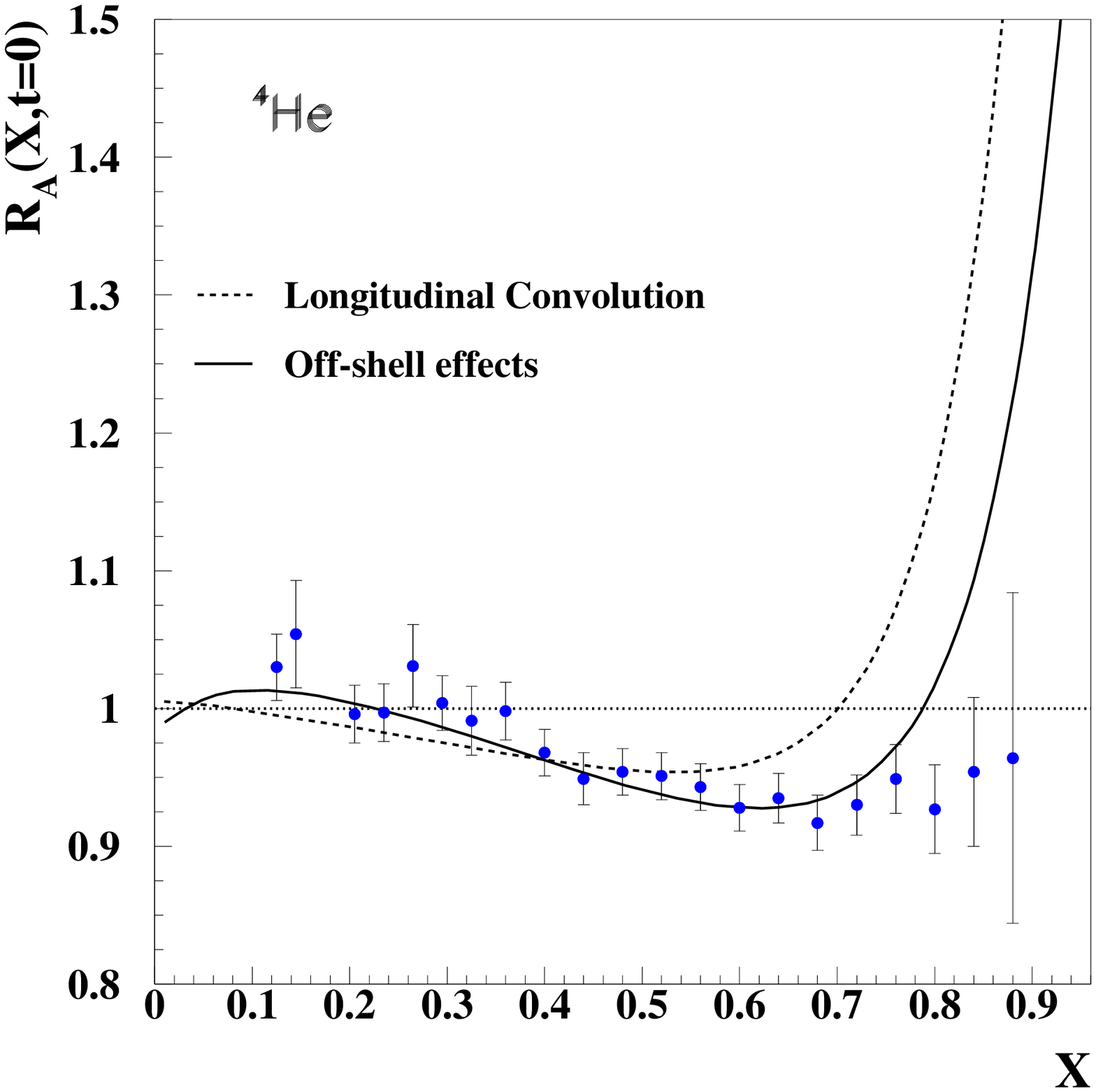}
\includegraphics[width=8.cm]{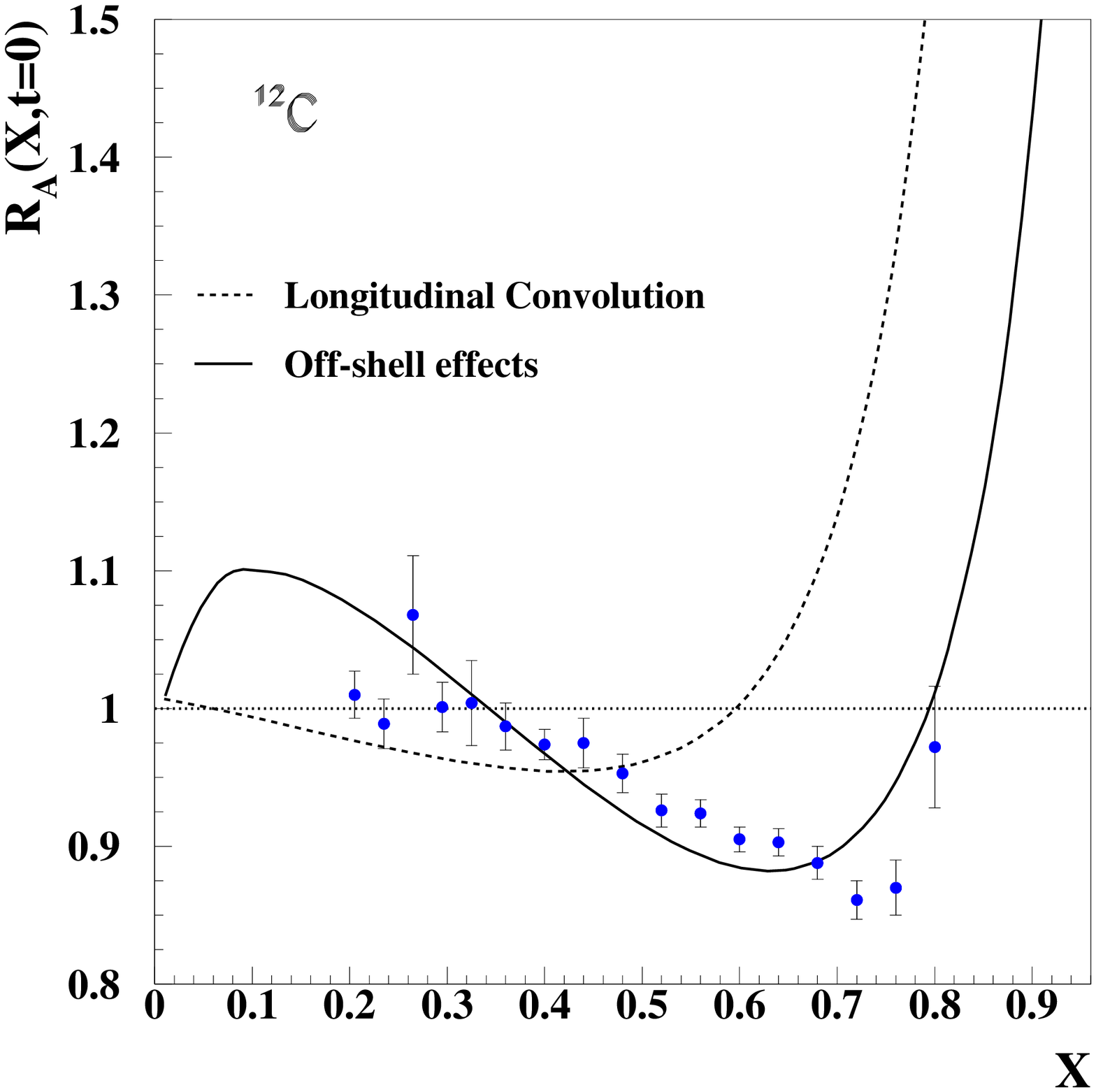}
\caption{EMC effect in forward Compton scattering, with off-shell effects (full line), 
Eq.(\protect\ref{HA}),
and calculated with the longitudinal convolution formula,
Eq.(\protect\ref{HA_convo}). Left panel: $^4$He, right panel $^{12}$C. 
Experimental data from \protect\cite{Gomez}. } 
\label{fig3}
\end{figure}

The main conclusion from these papers
was that nucleon off-shell effects are essential
for reproducing quantitatively the effect although
the dynamical source of off-shellness was not transparent. 
In Fig.\ref{fig3} we present a calculation of the ratio $R_A^{DIS}= d \sigma_A/d \sigma_D$ 
obtained including off-shell effects. 

In order to interpret the curves in the figure, a few comments are in order:

\noindent {\it i)} For consistency with the presentation of experimental data, 
we show ratios of forward nuclear
structure functions (with $A>2$),
to the deuteron forward structure function. The deuteron is in itself 
affected by an EMC effect \cite{AKL}. Ratios of $A>2$ nuclei to a nucleon 
structure function will in principle look different.

\noindent {\it ii)} The dashed curve is calculated according to 
the longitudinal convolution, {\it i.e.} with no off-shell effect included.
The effect of binding is enhanced with respect to mean field models
\cite{Mil_A} because of the account of short range correlations among 
nuclei \cite{CioLiu}. We note however that despite the differences between 
the models in \cite{Mil_A} and \cite{CioLiu}, neither one is able to 
reproduce the effect for the measured range of values of nuclei. 
In other words binding cannot account for the $A$ dependence of the effect.

\noindent {\it iii)} As already noted in \cite{KPW,GroLiu}, {\it kinematical} 
off-shell effects, coming from the different relation between the active 
quark's virtuality and its transverse momentum in a nucleus, 
by generating an extra $A$-dependent term, can account
for the discrepancies at large $x$. This is clearly seen 
from Fig.\ref{fig3}.         

\noindent {\it iv)} {\it Dynamical} off-shell effects, originating from 
intrinsic modifications of the quark spectral function need to be taken into 
account in order to guarantee the fulfillment of Adler's Sum Rule \cite{adler} 
in nuclei. Forward DIS, besides allowing us to state that dynamical 
off-shell effects should indeed be present, gives very little insight 
on the nature of the effects. Our calculation presented in Fig.\ref{fig3}, 
includes the effect of parton reinteractions, located mainly at 
low $x$ ($x<0.3$) \cite{Liu-new}. This seems to be a favored mechanism
with respect to the nucleon deformation one, taken into account 
in \cite{KPW}.

\noindent {\it v)} Most crucially,  
nucleon binding alone does not provide the correct $A$-dependence 
of the EMC effect ratio because of the saturation with in increasing $A$ 
of both the kinetic and separation energies of nucleons' 
\cite{CdALiu_Adep,Mil_A}. Off-shell effects provide an extra degree
of freedom that allows for a quantitative description of the effect.

GPDs represent a theoretical tool 
that can be used for studies of both parton reinteractions \cite{XWang}, and
of nucleons spatial dimensions \cite{Bur,Diehl03,LiuTan1}. 

It is therefore natural to extend our approach
to the off-forward case. We discuss the region: $\zeta \leq X$, which is most relevant
for the EMC effect, where both partons in
Fig.\ref{fig1} are quarks (an extension to the region $\zeta > X$, where GPDs behave
like meson distributions, will be considered in a forthcoming paper). 
 
\subsection{Nucleon}
\label{nucleon-sec}
In the spectator model for the nucleon \cite{LiuTan1},
the outgoing diquark has spin/isospin either $0$ or $1$. 
Its mass is in principle described by a 
mass distribution or spectral function.
In our calculations we consider only the scalar component of the
diquark. Furthermore, we consider the mass spectrum to be dominated
by a single mass value. 
Both assumptions can in principle be lifted at a later stage.
However, for our goal of studying nuclear effects, 
we do not expect such details of the model to produce 
significant changes. 
The structure of the nucleon matrix element can be read from the upper 
part of 
Fig.\ref{fig2}:
\begin{eqnarray}
{\cal M}^N_{ij} = 
\overline{U}(P^\prime,S) \overline{\Gamma}(k^\prime,P) 
\, \frac{(\not\!k^\prime+m)}{k^{\prime 2}-m^2} 
\, \frac{({\not\!k} + m)}{k^2-m^2} \Gamma(k,P) U(P,S), 
\label{model_N1}
\end{eqnarray}
where $\Gamma$ is the vertex connecting the quark, diquark, and nucleon;
$U(P,S)$ is the nucleon spinor.
The form of $\Gamma$ depends on the 
way the spin and isospin are carried through the vertex.
In analogy with the forward case \cite{GroLiu}, we define $\Gamma$ so that:
\footnote{Similar results were obtained in \cite{JakMul}
by assuming  $\Gamma= \hat{I} \Phi(k^2,P^2)$. 
The complete set of independent functions 
that appear in principle using a covariant formalism for 
the off-shell tensor was listed in \cite{MelSchTho}.}
\begin{eqnarray}
{\cal M}^N_{ij}  =  \rho_N \left( k^2, k^{\prime \, 2} \right) \, 
\sum_s u_i(k,s) 
\overline{u}_j(k^\prime,s) 2 \pi \delta(k_X^2 -M_X^2), 
\label{matrix2}
\end{eqnarray}
with 
\begin{equation}
\rho_N \left( k^2, k^{\prime \, 2} \right) = 
{\cal N} \frac{\phi(k^{\prime\, 2})}{k^{\prime 2}-m^2} \frac{\phi(k^2)}{k^2-m^2},
\label{rho}
\end{equation}
${\cal N}$ being a normalization constant.
In Eq.(\ref{matrix2}) 
the active off-shell parton is described
by a spinor that obeys the Dirac equation, with an off-shell mass, $m^*$
($m^* = (m + \sqrt{k^2})/2$):
\begin{equation}
(m + \not\!k) u(k,s) = 2 m^* u(k,s)
\label{dirac}
\end{equation}
By inserting Eqs.(\ref{model_N1}) and (\ref{matrix2}) in 
Eqs.(\ref{h_tensor}) and (\ref{sfn})
one obtains: 
\begin{eqnarray} \sqrt{1-\zeta} H^N(X,\zeta,t) -
\frac{1}{4}\frac{\zeta^2}{\sqrt{1-\zeta}}
E^N(X,\zeta,t)
&  =  & \nonumber \\  =  \frac{1}{2P^+} \int \frac{d^4 k}{(2\pi)^4} \, 
{\rm Tr} \left[ \gamma^+ {\cal M}^N(k,P,\Delta) \right]  = & & 
\nonumber \\
 =  \frac{X}{1-X} \sqrt{\frac{X-\zeta}{X} }
\int \frac{d^2 k_{\perp}}{(2\pi)^3} \, \rho_{N}({k^2},k^{\prime \,2}). & &
\label{HN}
\end{eqnarray}

The square root factor in the equation takes into account the normalization 
of the spinors from Eq.(\ref{dirac}).  
In our derivation we made use of Eqs.(\ref{Fq}); all kinematical variables have been 
defined according to Eqs.(\ref{kin2}) with $A=1$. More details are given
in the Appendix.

The results obtained above correctly reproduce the parton model 
in the forward limit where:
\begin{eqnarray}
\overline{U}(P,S) \overline{\Gamma}(k,P) 
\frac{(\not\!k+m)}{k^{\prime 2}-m^2} 
\frac{({\not\!k} + m)}{k^2-m^2} \Gamma(k,P) U(P,S) & \approx & \rho_N(k^2) 
(\not\!k+m), 
\label{model_N3}
\end{eqnarray}
yielding for the (transverse) structure function:
\begin{eqnarray}
F^N(x_{Bj}) & = & \frac{1}{2P^+} \int \frac{d^4 k}{(2 \pi)^4}  \, \rho_N(k^2)  \, {\rm Tr} \left[ \gamma^+ (\not\!k+m) \right] \nonumber \\
& = & \frac{x_{Bj}}{1-x_{Bj}} \int \frac{d^2 k_\perp}{(2\pi)^3} \rho_N(x,k_\perp^2)
\label{sfn_forward}
\end{eqnarray}
with $\rho_N(k^2,k^{\prime 2}) \rightarrow \rho_N(k^2)$ for $k=k^\prime$. 


\subsection{Nucleus}
The nuclear part of the diagram in Fig.\ref{fig2} is evaluated
similarly to the nucleon one. 
For a spin $0$ nucleus 
we write the matrix element, Eq.(\ref{A_tensor}), as: 
\begin{eqnarray}
{\cal M}^A_{ij} = 
\overline{U}_{A-1}(P_A^\prime,S) \overline{\Gamma}_A \, 
\frac{({\not\!P^\prime} + M)}{P^{\prime 2}-M^2}  \, \frac{({\not\!P} + M)}{P^2-M^2} \Gamma_A \, U_{A-1}(P_A,S) 2 \pi \delta(P_{A-1}^2-M_{A-1}^{* \, 2}) 
\label{model_A1}
\end{eqnarray}
where $U_{A-1}$ describes the $A-1$ system, and $\Gamma_A$ is the nuclear 
vertex function, and $M_{A-1}^{*}$ is the mass of the outgoing $A-1$ nuclear system. 

We now make the assumption that the spin structure 
at the nuclear vertex is treated similarly 
to Eq.(\ref{matrix2}):
\begin{eqnarray}
{\cal M}^A_{ij} = \rho_A(P^2,P^{\prime 2}) \, \sum_S 
U_i(P,S) \overline{U}_j(P^\prime,S), 
\label{matrixA}
\end{eqnarray}
$U_i(P,S)$ being the
spinor for an off-shell nucleon with effective mass $M^* \neq M$. 
$\rho_{A}(P^2,P^{\prime \, 2})$ is the off-diagonal
nucleon light cone momentum distribution. It can be approximated by a 
non relativistic nuclear spectral function:
\begin{eqnarray}
\rho_A(P^2,P^{\prime \, 2}) & \approx & 
S_A(\mid {\bf P} \mid ,\mid {\bf P^\prime}\mid,E) \nonumber 
\\
& = & \sum_f \Phi_f(\mid {\bf P} \mid ) \Phi_f^*(\mid {\bf P^\prime} \mid)
\delta\left( E-(E_{A-1}^f -E_A) \right),  
\label{rho_A}
\end{eqnarray} 
where $\mid {\bf P} \mid$, and $\mid {\bf P}^\prime \mid$ are the 
absolute values of the incoming and outgoing nucleons three-momenta, 
respectively, $E$ is the nucleon separation energy, $E_A$ being the 
binding energies of the initial nucleus, $A$, and of the final nuclear
system, $A-1$. $\Phi_f$ is the (Fourier transformed) overlap integral between 
the initial and final nuclear wave functions; the sum over $f$ is carried out 
over all the final configurations of the $A-1$ system (see \cite{CioLiu} and references
therein for more details). 
Eqs.(\ref{matrixA}) and (\ref{HN}), inserted in  (Eq.(\ref{convo1})), 
yield:
\begin{eqnarray}
F^{A}(X,\zeta,t)= \int \frac{d^2 P_{\perp}{dY}}{2(2\pi)^3}
 \, \frac{1}{(A-Y)}  {\cal A} \rho_{A}(P^2,P^{\prime \, 2}) 
\, {F^{N}_{OFF}(X_N,\zeta_N,P^2,t)},
\label{FA}
\end{eqnarray}
where ${\cal A} = (Y-\zeta/2)(\sqrt{Y(Y-\zeta)})$ accounts for 
the spinors normalization (Eq.(\ref{matrixA})). 
By replacing $F^N_{OFF}$ with the expression in Eq.(\ref{HN}), one finally obtains: 
\begin{eqnarray}
H^A(X,\zeta,t)  =  \int \frac{d^2 P_\perp dY}{2(2\pi)^3}
\frac{1}{(A-Y)}  \rho_A 
\left[ P^2(Y,P_\perp^2),P^{\prime \, 2}(Y,P_\perp^2,\zeta,t) \right] 
{\cal A} & & \\
\times  
\sqrt{\frac{Y-\zeta}{Y}} \left[H^N_{OFF}\left(\frac{X}{Y},\frac{\zeta}{Y},P^2,t \right) -
\frac{1}{4} \frac{\left(\zeta/Y \right)^2}{1-\zeta/Y} 
\, E^N_{OFF}\left( \frac{X}{Y},\frac{\zeta}{Y},P^2,t \right) \right] & &
\label{HA}
\end{eqnarray}
The kinematical variables 
are from Eqs.(\ref{kin2});
similarly to the free nucleon case.
Numerical results presented in the following where obtained in the
small $\zeta$ approximation, {\it i.e.} considering 
for the off-shell nucleon:
\begin{eqnarray}
\sqrt{1-\zeta_N} H^{N}_{OFF}(X_N,\zeta_N,P^2,t)  =  \frac{X_N}{1-X_N}
\sqrt{\frac{X_N-\zeta_N}{X_N}} \int \frac{d^2 k_\perp}{2 \pi} \,
\widetilde{\rho}_{N}\left[ k^2(P^2), k^{\prime \, 2}(P^2) \right],
\label{bound_F}
\end{eqnarray}
and disregarding the second term in Eq.(\ref{HA}). $H^{N}_{OFF}$ is  
modified both kinematically and dynamically with respect to
the free nucleon one, $H^N$ (Eq.(\ref{HN})). 
Kinematical modifications 
due to Fermi motion and nuclear binding produce an extra shift
in the $X$ dependence with respect to the free nucleon one, analogous 
to what found in forward DIS. 
$H^{N}_{OFF}$ is however expected to be also 
structurally different from the on-shell case because of nuclear
medium induced distortions (schematically, $\widetilde{\rho} \neq \rho$).   
Off-shell modifications, differently from Fermi motion and binding,
affect the transverse variables. 
It is therefore of the out-most 
importance to evaluate them using GPDs, with a twofold goal in mind:
On one side because of the interpretation of GPDs as
Fourier transforms of Impact Parameter dependent 
Parton Distribution Functions (IPPDFs), such studies will provide a handle to 
directly evaluate the spatial modifications of the nucleon inside
the nuclear medium \cite{LiuTan3}.   
The other compelling reason to explore quantitatively  
off-shell effects in this context is that they are essential 
in order to interpret the experimental data on DIS from nuclei
(see Fig.\ref{fig3}). 

Kinematical off-shell effects can be evaluated straightforwardly as
they result from a difference in the 
relationship between the struck parton's virtuality, $k_\mu^2 \equiv k^2$, and
its intrinsic transverse momentum, $k_\perp$ in a free 
and in a bound nucleon, respectively:
\begin{subequations}
\begin{eqnarray}
k^2_N & = & {\cal M}^2(X) - \frac{{\bf k}_\perp^2}{1-X}  \\
k^2_A & = & {\cal M}_A^2(X_N) - \frac{{\bf a}_\perp^2}{1-X_N}, 
\end{eqnarray}
\end{subequations}
with:  
\begin{subequations}
\begin{eqnarray}
{\cal M}^2(X) & = & X M^2 + \frac{X}{1-X} M_X^2  \\
{\cal M}_A^2(X_N) & = & X_N P^2 - \frac{X_N}{1-X_N} M_X^2   \\
P^2 & = &\frac{Y}{A} \left(M_A^2 - \frac{M^2_{A-1} + P_\perp^2}{1-Y/A} - \frac{P_\perp^2}{Y/A} \right) . 
\end{eqnarray}
\end{subequations}
where $X_N=X/Y$, $m_{qq}$ is the diquark's mass, and the intrinsic transverse momentum in the nucleus 
is: ${\bf k}_\perp - X_N{\bf p}_\perp \equiv {\bf a}_\perp$. 

In a nucleus one has therefore a shift in the longitudinal variable, $X \rightarrow X_N$, a shift in the
transverse variable, ${\bf k}_\perp \rightarrow {\bf k}_\perp - X_N{\bf p}_\perp$, and a shift in the 
nucleon's invariant mass squared, $M^2 \rightarrow P_\mu^2$.  
The shift in $X$ is associated with the $X$-rescaling caused by binding. 
The shifts in the transverse variables are due to off-shell effects. 
Even in the absence of nucleon deformations ($\widetilde{\rho} =\rho$), these shifts modify 
the bound GPD, $H^{N}_{OFF}$ in Eq.(\ref{HA})).
It is important to notice that such modification is based on the same
physical mechanism as for the bound forward structure functions. By changing variables 
from $k_\perp$ to $k^2_\mu$, and by writing explicitly $\rho$ as in Eq.(\ref{rho}), one can in fact 
write:     
\begin{eqnarray}
H^{N}_{OFF}(X_N,\zeta,P^2,t) &  =  & X_N
\sqrt{\frac{X_N}{\zeta -X_N}} \int_0^{2 \pi} d \phi \int^{k_{A \, MAX}^2(X_N)} \frac{d k_A^2}{2 \pi} \,
\frac{\phi(k_A^2) \phi^*(k_A^{\prime \, 2})}{ k^2_A k_A^{\prime \, 2}}, \\
H^{N}(X,\zeta,t) &  =  & X
\sqrt{\frac{X}{\zeta -X}} \int_0^{2 \pi} d \phi \int^{k_{N \, MAX}^2(X)} \frac{d k_N^2}{2 \pi} \,
\frac{\phi(k_N^2) \phi^*(k_N^{\prime \, 2})}{ k^2_N k_N^{\prime \, 2}}, \\
\label{bound_F_2}
\end{eqnarray}
The difference between $H^{N}_{OFF}$ and $H^{N}$ is in the upper limit of integration for the equations
above. 
The effect of kinematical off-shellness  
can therefore be obtained as an
additional rescaling of the longitudinal variable, $X$, that turns out to increase the 
effect of Fermi motion and binding. 
Phrased otherwise, 
kinematical off-shell effects are an indirect 
consequence of Fermi motion and binding that, although originating from modifications of 
transverse variables in a nucleus, affect the dependence of the 
GPDs on the longitudinal variable $X$. 

The existence of kinematical off-shellness
indirectly implies that intrinsic deformations/parton reinteractions are present.
In other words, off-shell effects are an indirect manifestation of the 
impact of interactions
among particles during the hard scattering process. A clear illustration
of this picture was given, for example, in \cite{EFP} where 
it was shown that in DIS from a nucleon kinematical off-shell effects
can generate $F_L$; however, this can be evaluated consistently
with baryon number conservation (Adler's sum rule \cite{adler}) 
only by including {\it ab initio} interactions, in this case 
the higher twist terms. 
In the nuclear case a full theory of parton re-interactions is still 
lacking (see however many aspects treated {\it e.g.} in \cite{XWang} and 
references therein), 
therefore in what follows 
we introduce a phenomenological approach.
We first calculate exactly the kinematical off-shell effects. We 
then restore Adler Sum Rule by introducing interactions through a 
modification of spectral function's denominator, Eq.(\ref{rho}). 
A more detailed discussion of this point is beyond the scope
of this paper and will be given elsewhere \cite{Liu-new}.
 
%
%
If instead off-shell effects are disregarded, one recovers a 
longitudinal ``convolution formula'':
\begin{equation}
H^A(X,\zeta,t)= \int_X^A dY \, \sqrt{\frac{Y-\zeta}{Y}} \, f_A(Y,\zeta,t) \, 
H^N(X_N,\zeta,t),
\label{HA_convo}
\end{equation}
where for consistency with previous literature in forward scattering 
\cite{Tho}, in Eq.(\ref{HA}) we switch integration variables from $P_\perp$ to 
$\mid P \mid \equiv P$:
\begin{equation}
\label{fz}
f_A(Y,\zeta,t)  =  2 \pi M \int_{P_{min}(Y,\overline{E})^\infty} d P P \Phi_A(P) 
\Phi_A^*(\mid {\bf P} + {\bf \Delta} \mid).
\end{equation}
Other kinematical variables that appear in the equation are:   
\begin{subequations}
\begin{eqnarray}
\mid {\bf P} + {\bf \Delta} \mid & = & (P^2 + \zeta^2 P_{\|}^2 + \Delta_\perp^2 + 2 P_{\|}\zeta + 2 P_\perp \cdot \Delta_\perp)^2 \\
P_{\|} & \approx & (M - \overline{E}) - M Y  \\
P_{min}(Y,\overline{E}) & = & \mid M(1-Y) - \overline{E} \mid.
\end{eqnarray}
\end{subequations}
$\displaystyle \overline{E} = \int d^3 P dE S_A(P,E) E$, is the average separation energy. 
Its value, enhanced by nucleon short range correlations, was shown to govern
binding effects in the forward EMC effect (see \cite{Tho} and references therein).  
By performing the integration  
over 
$\displaystyle \mid {\bf P} \mid \equiv P = (P_{\|}^2 + P_\perp^2)^{1/2}$,
instead than over the 
transverse momentum $P_\perp$, 
one generates a $Y$-dependent 
lower limit of integration, $P_{min}$. 
The expression for $P_{\|}$ was obtained by
energy-momentum conservation at the nuclear vertex in Fig.\ref{fig2}, disregarding the recoil
energy of the $A-1$ nuclear system. 

We conclude this subsection by noting that in the forward limit, $\Delta \rightarrow 0$,
Eqs.(\ref{HA},\ref{HA_convo}) correctly reproduce the nuclear formulae obtained {\it e.g.} 
in Refs.\cite{GroLiu,CioLiu}.
 
\subsection{Numerical Results}
We calculated $H^A$ in both the on-shell and off-shell cases using the nuclear
model from Ref.\cite{CioLiu} and a parametrization for $H^N$ at $\zeta=0$
from Ref.\cite{LiuTan1}. All results are presented for the $^4He$ nucleus, 
although our formalism applies more generally to complex nuclei with larger $A$.   
\footnote{Calculations for other complex nuclei are available under request at 
the e-mail addresses listed here.}
\subsubsection{Off-forward EMC Effect}
\begin{figure}
\includegraphics[width=16.cm]{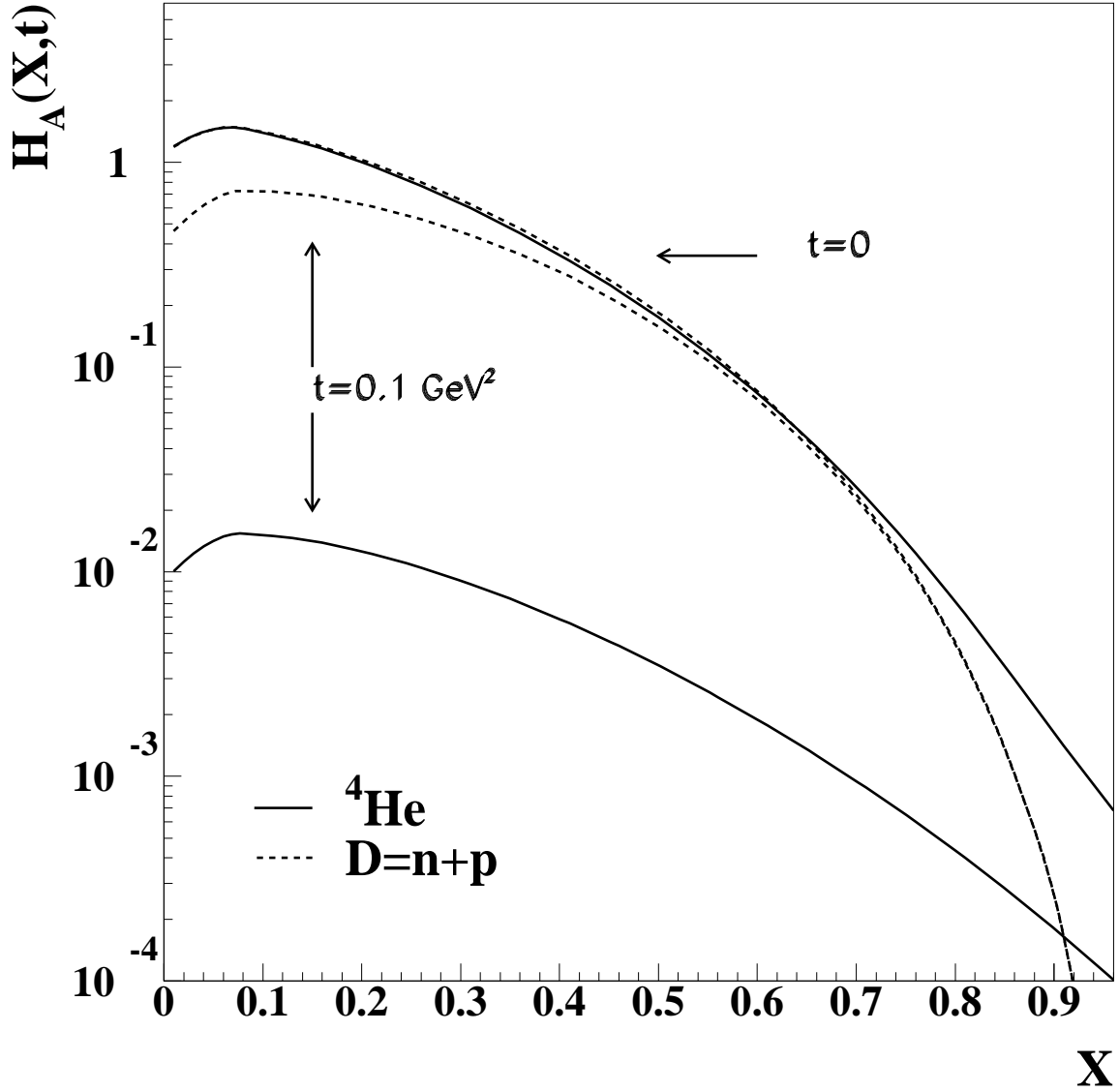}
\caption{GPD in $^4He$, divided by the number of nucleons, (full lines) and in the nucleon (dashed lines), evaluated at
two different values of $t$, $t=0$, and $t=0.1$ GeV$^2$, as respectively indicated by
the arrows in the figure. 
} 
\label{fig4}
\end{figure}

\begin{figure}
\includegraphics[width=16.cm]{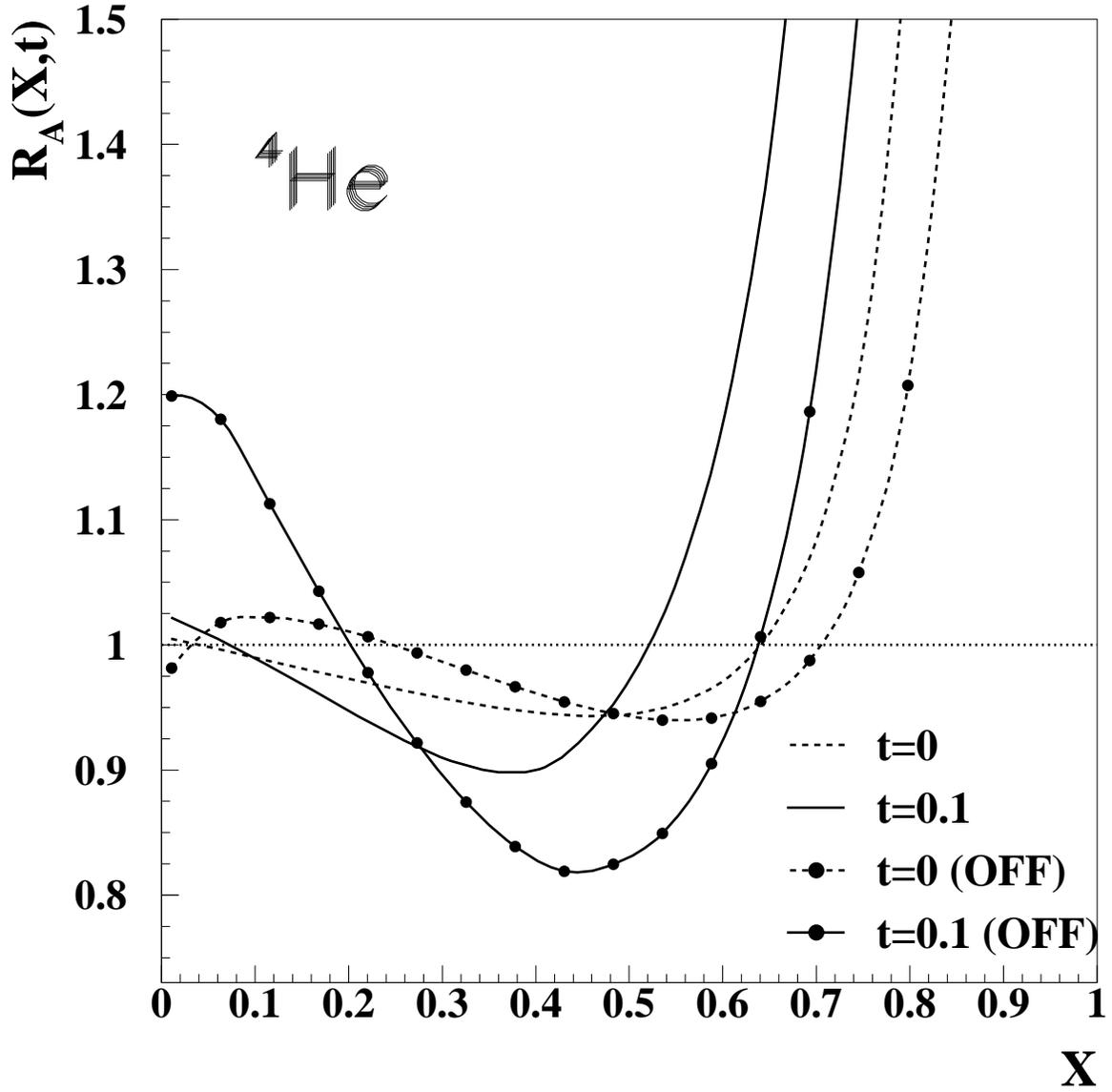}
\caption{Off-forward EMC effect in $4$He. The ratio $R^A$ represented in the figure is defined in 
Eq.(\ref{RA}). Full line: including off-shell effects. 
Dashed line: longitudinal convolution. The full line with bullets includes off-shell
effects, at $t=0.1$ GeV$^2$; the dashed line with bullets was obtained with the longitudinal
convolution at $t=0.1$ GeV$^2$. The lines without bullets correspond to $t=0$.} 
\label{fig5}
\end{figure}

In Fig.\ref{fig4} we show the $H^{N(A)}(X,0,t)$ plotted vs. $X$,
for a nucleus ($^4He$) and for the proton, both normalized to $1$, 
and for $t=0$ and $t=0.1$ GeV$^2$, respectively.
Because $^4$He has a much larger drop with $t$ due to the nuclear form factor
behavior, we consider the ratio:
\begin{equation}
R^A(X,\zeta=0,t) = \frac{H^A(X,t)/F^A(t)}{H^N(X,t)/F^N(t)} 
\label{RA}
\end{equation}
where: $H^{N(A)}(X,t) \equiv H^{N(A)}(X,\zeta=0,t)$, and $F^N$ is the Dirac form factor
for a proton. 
$R^A$ becomes equal to the EMC ratio in the forward limit $t=0$. The following normalizations 
hold: 
\begin{equation}
\int dX \frac{H^{N(A)}(X,t)}{F^{N(A)}(t)} = 1. 
\end{equation} 
By choosing the form for $R^A$ in Eq.(\ref{RA}), one eliminates the, somewhat 
spurious, $t$ dependence coming from the mere comparison of form 
factors of nuclei with different $A$ (see Fig.\ref{fig4}). 
Nevertheless, $R^A$ includes contributions from both the long range and short
range nuclear structure. 
In Fig.\ref{fig5} we show $R^A$ as a function of $X$ for different values 
of $t$ ($t= 0$ GeV$^2$, and $t=0.5$ GeV$^2$), both including (Eq.(\ref{HA})), and disregarding (Eq.(\ref{HA_convo}))
off-shell effects. Notice that the curves for $t=0$ do not coincide exactly with the ``EMC''-like ones
in Fig.\ref{fig3}, because they are divided by the proton, while 
Fig.\ref{fig3} includes nuclear effects in the deuteron target.
The EMC-effect is both enhanced in magnitude,
and shifted to lower values of $X$, at $t \neq 0$. Furthermore the effects of dynamical off-shellness are 
enhanced. As we explain also below, it should be clarified that such an enhancement is not a direct consequence of
having probed through GPDs partonic and/or nucleonic transverse degrees of freedom. In fact, because $\Delta$ and 
the targets intrinsic transverse momenta are decoupled, the $t$ dependence in $R^A$ originates from the modification
of the longitudinal variable: $X \rightarrow X/\langle Y(A,t) \rangle$, the average value of $Y$ being calculated 
using Eq.(\ref{rho_A}).    

\begin{figure}
\vspace{3cm}
\includegraphics[width=8.0cm]{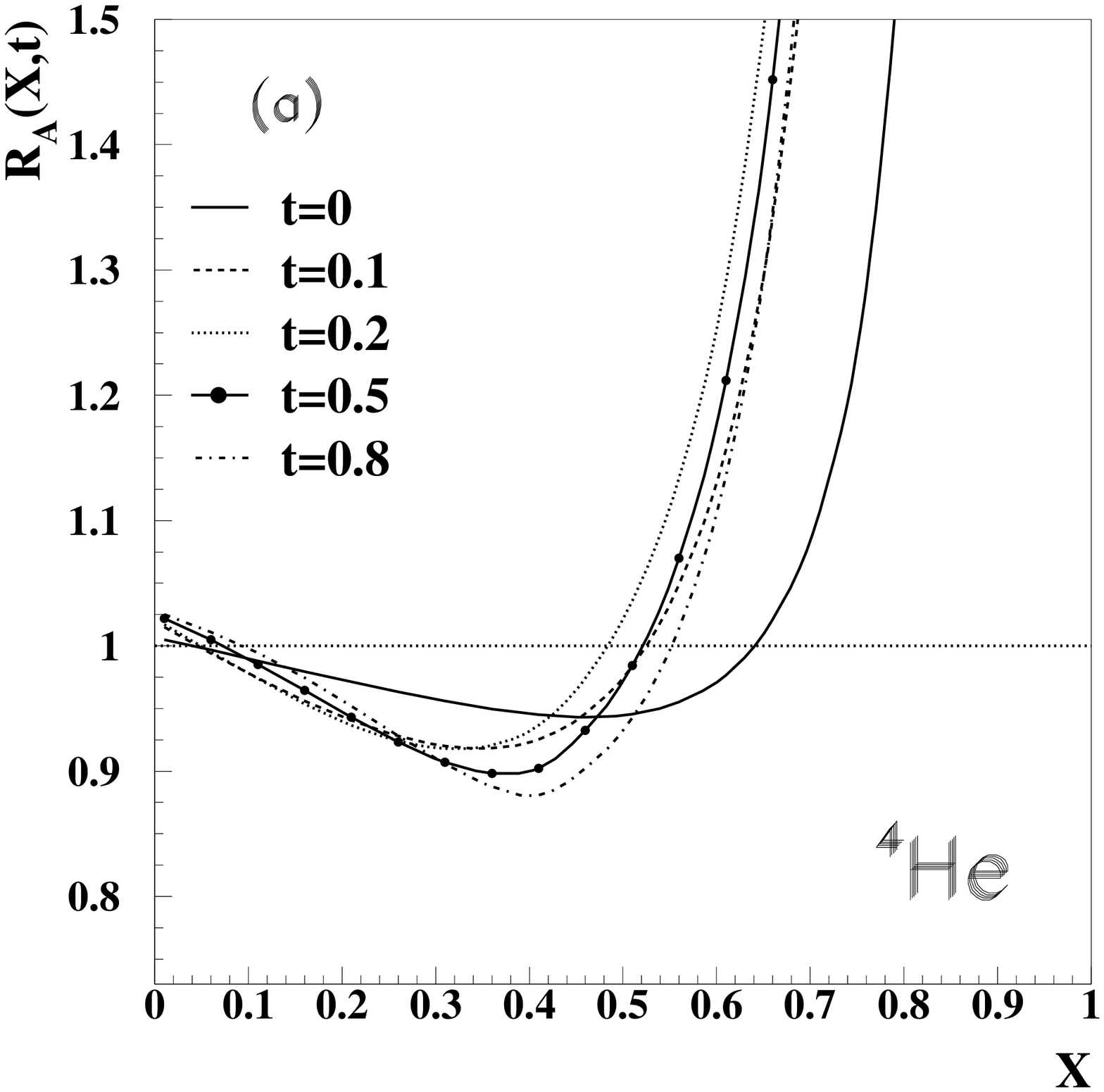}
\includegraphics[width=8.0cm]{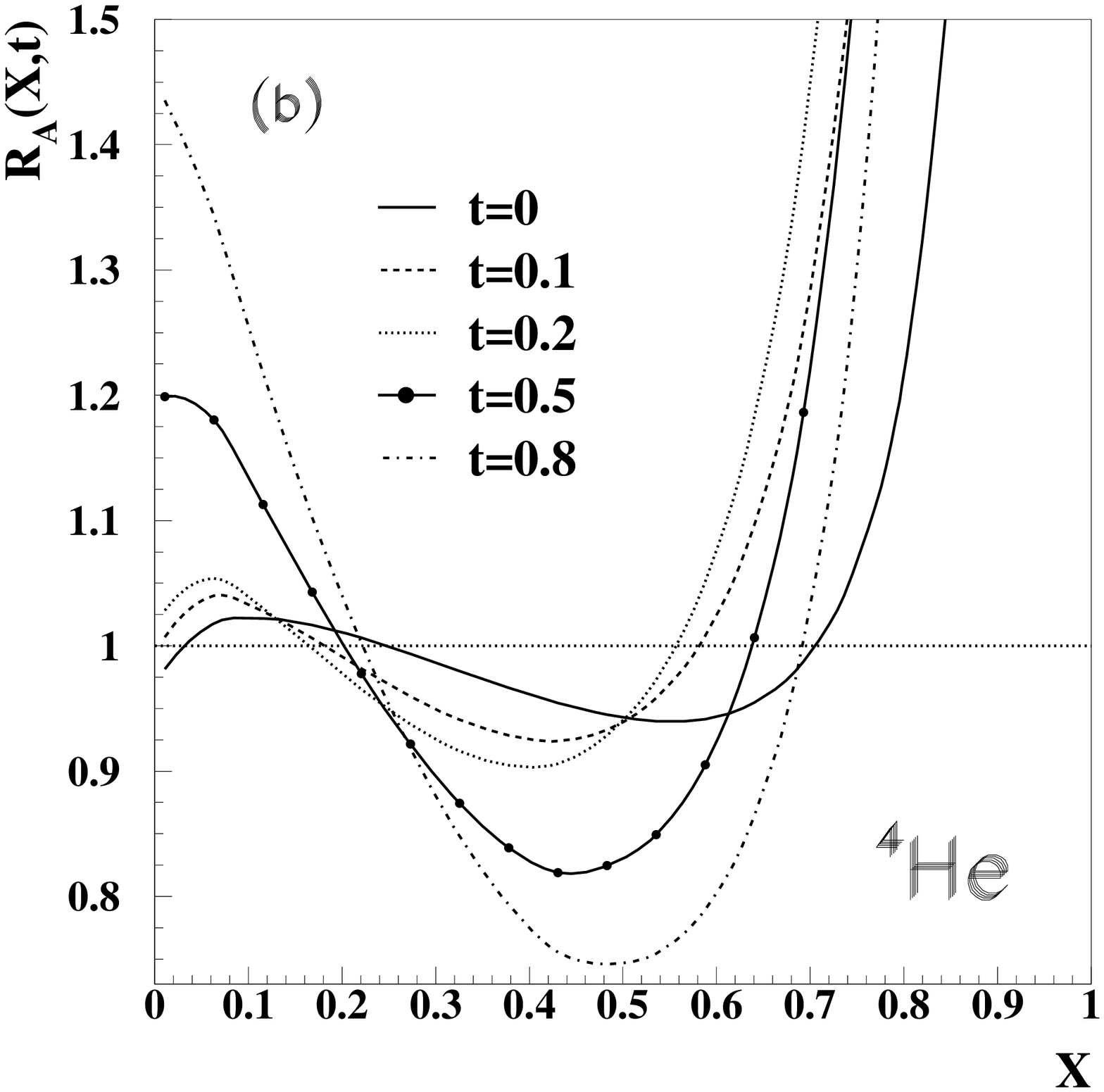}
\caption{Comparison of off-forward EMC effect without 
(panel {\bf (a)}), and with (panel {\bf (b)}) off-shell effects. The ratio $R^A$ is given 
in Eq.(\ref{RA}). Several
values of $t$ ranging from $t=0$ to $0.8$ GeV$^2$ are considered, as labeled in the figure.} 
\label{fig6}
\end{figure}

In Fig.\ref{fig6}, we show the $X$ behavior of $R^A$ for different values of $t$, in separate
panels: longitudinal convolution (left), off-shell effects (right).    
The marked difference at $x=0.1$ was unexpected according to estimates based 
on the longitudinal convolution \cite{KirMul}. The effect predicted in this paper is quite sizable and it
should be measurable within the accuracy of data 
currently been analyzed at HERMES. 

\subsubsection{$t$-dependence and Role of Long Range Nuclear Interactions}
In order to proceed further, one must devise a method to 
disentangle the different contributions from both the short and long range nuclear effects originating from the $t$ dependent function $\rho_A$ in Eq.(\ref{HA}).
A number of long range effects that are well known since the early evaluations of 
nuclear form factors,  are also present in principle for the nuclear GPDs.
In a nutshell, one must include 
terms beyond the nuclear impulse approximation -- the so-called two-body and 
three-body
current contributions from the coupling of the virtual photon 
with a mesonic component
in the nucleus, and with a nucleon resonance, respectively \cite{KirMul}.    
\vspace{0.3cm}

{\em To what extent are these terms affecting the ratio $R^A$?}

\vspace{0.3cm}
\noindent
In order to evaluate their separate contributions, we consider a Taylor 
expansion of Eq.(\ref{RA}) 
around the value $Y=1$, and $P^2=M^2$:
\begin{eqnarray}
\displaystyle
R^{A}(X,t) & \approx & 1 + 
\frac{\langle Y_1^A(t) \rangle}{F^{A, point}(t)}  \times
\left[ \frac{1}{H^N(X)} \frac{\partial H^N(X,t)}{\partial X} X \right]  \nonumber \\ 
& + &  \frac{\langle Y_2^A(t) \rangle}{F^{A, point}(t)}  \times
\left[X^2\frac{ \frac{\partial^2 H^N(X,t)}{\partial X^2} -
2X \frac{\partial H^N(X,t)}{\partial X} }{2 H^N(X)}\right] ..., 
\label{Taylor1} 
\end{eqnarray}
where, from Eqs.(\ref{HA_convo}) and (\ref{1moment}), 
$F^A(t) = F^{A, point}(t) F^N(t)$, with:
\begin{eqnarray}
F^{A, point}(t) & = & \int_0^A dY \, \rho_A(Y,t) \\     
\langle Y_1(A,t) \rangle & = & \int_0^A dY \, \rho_A(Y,t) (1-Y)  \\ 
\langle Y_2(A,t) \rangle & = & \int_0^A dY \, \rho_A(Y,t) (1-Y)^2, 
\end{eqnarray}
$F^{A, point}$ is the nuclear form factor, calculated by assuming point-like 
nucleon structure. $Y_i$, $i=1,2$ are proportional to higher nuclear 
moments. 
Eq.(\ref{Taylor1} helps us understand the behavior in $X$, $A$, and $t$ of $R^A$, 
from the observation that such ratio
is determined by a combination of the values of 
$\displaystyle \langle Y_{1(2)}^A(t) \rangle \approx \langle Y_{1(2)}(A,0) \rangle $,
and the slope of $H^N$ vs. $X$.     
This can be seen more intuitively by taking a cruder approximation for the numerator of $R^A$:
\begin{equation}
\frac{H^A(X,t)}{F^A(t)} \approx \frac{1}{F^N(t)} H^N\left( \frac{X}{\langle Y^{norm}(A,t) \rangle},t \right),   
\end{equation}
where $Y^{norm}(A,t)$ is the $t$-dependent average nucleon momentum in a nucleus, normalized to unity. The extent
to which $H^A$ on the l.h.s differs from the free  nucleon case is governed by the rescaling in $X$ shown
on the r.h.s.. This, in turn, depends both on the size of the deviation from unity of the nucleon momentum 
$\langle Y^{norm} \rangle$, and on the steepness of the slope of $H^N$ in $X$.    

\begin{figure}
\includegraphics[width=8.cm]{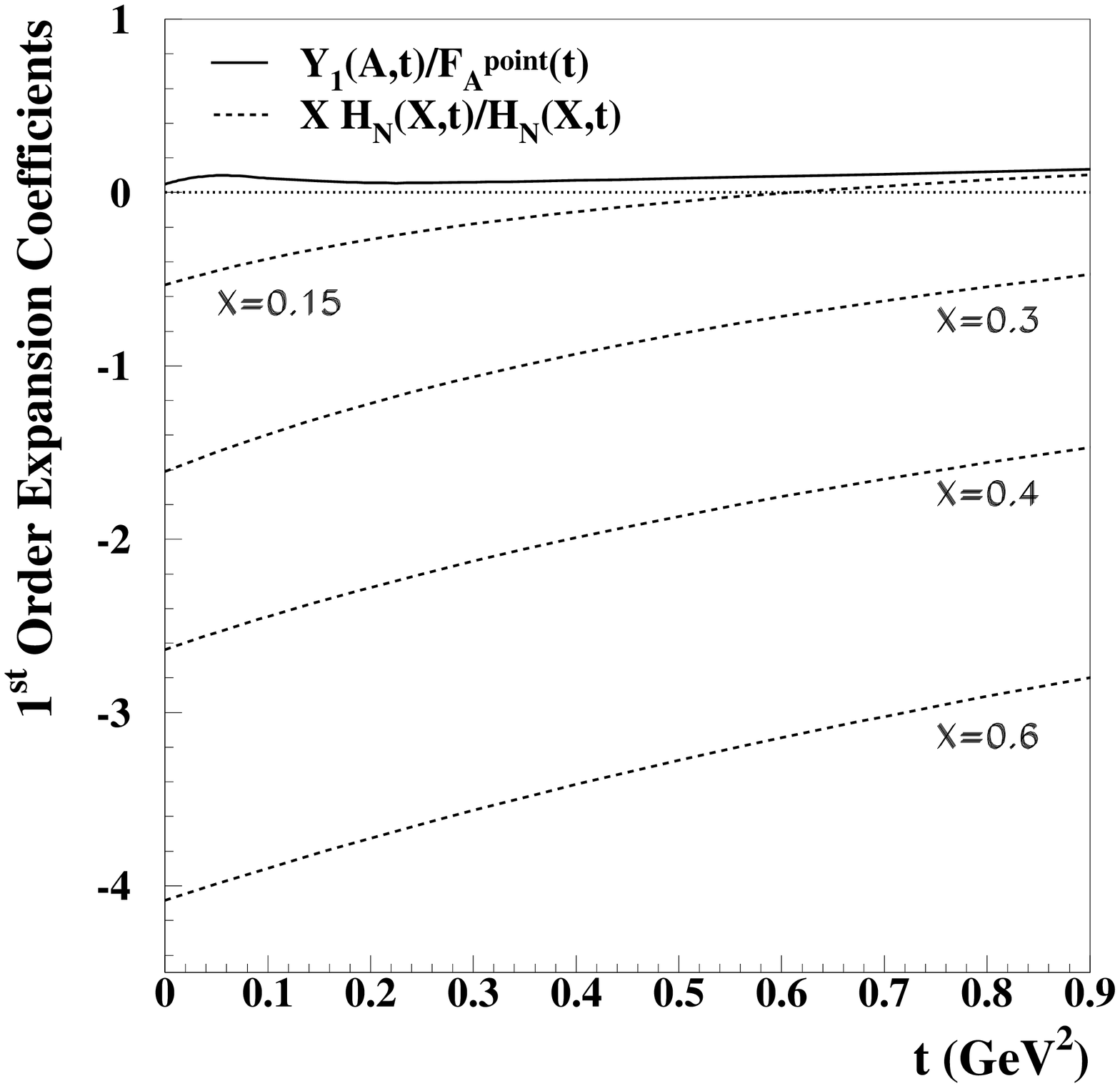}
\includegraphics[width=8.cm]{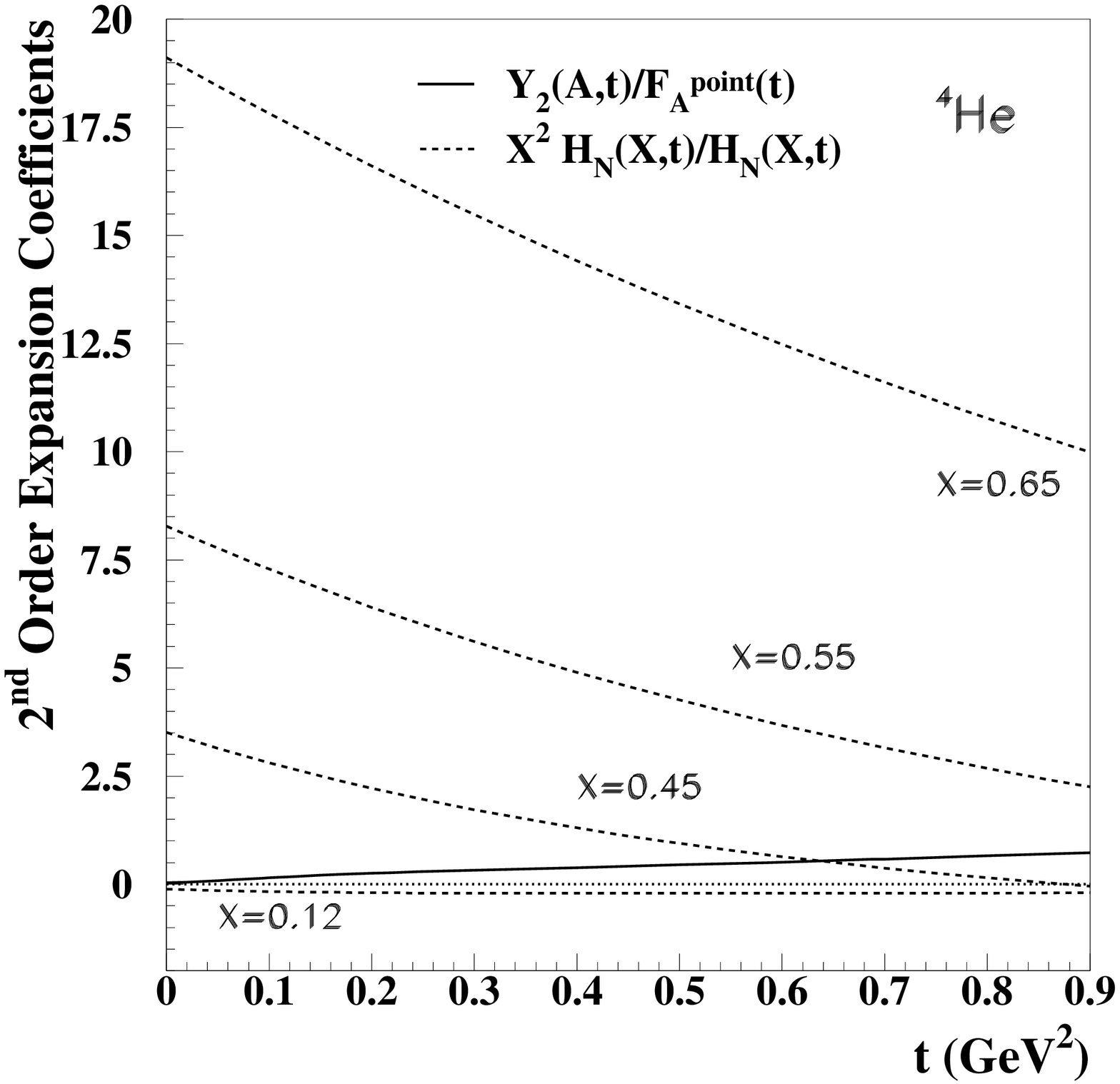}
\caption{Comparison between long range nuclear contributions, 
and deep inelastic type contributions the to the $t$ dependence of the ratio 
$R^A$, Eq.(\ref{RA}). 
By performing an expansion in the parameter $Y$, Eq.(\ref{Taylor1}), one can pinpoint
the driving terms for such dependence. The figure illustrates the first two
terms of the expansion, from which the nuclear ``form-factor-like'' dependence 
is clearly shown not to affect the ratio.} 
\label{fig7}
\end{figure}

In the forward limit, 
$Y_1(A,t=0) = \overline{E}_A/M - (1/3)\langle P^2 \rangle_A/M^2$,  
$Y_2(A,t=0) = (1/3)\langle P^2 \rangle_A/M^2$\cite{CioLiu}.   
Notice that all the higher order terms in the series 
have a similar structure, namely they are products of a 
$\langle Y_n(A,t) \rangle$, $n>1$ term, and a $n^{th}$ order derivative
of $H^N$ divided by $H^N$.
The terms: $\displaystyle \langle Y_{1(2)}(A,t) \rangle/F^{A, point}(t)$, and 
$\displaystyle (X \partial H^N(X,t)/\partial X)/H^N(X,t)$, 
$\displaystyle (X^2 \partial^2 H^N(X,t)/\partial^2 X^2)/H^N(X,t)$ are plotted in Fig.\ref{fig7}
vs. $t$, from which we see that the $t$ dependence coming from the 
$\displaystyle \langle Y_{1(2)}(A,t) \rangle$ terms is relatively very small. 

The $t$ dependence of $R^A$ shown  
Figs.\ref{fig5} and \ref{fig6} 
can be understood, therefore within the nuclear impulse approximation.
The mechanism that produces the modifications of the bound nucleon GPD, and the
predicted increase of the effect at $t \neq 0$, is, similarly to the forward
case, a reduction of the quarks momentum inside the nuclear medium. Such
reduction as we will show in more detail in Section \ref{moments-sec}, is enhanced 
at $t \neq 0$.    
As a final remark, we have explained the off-forward EMC effect using the longitudinal 
convolution approach while being however aware (see Fig.\ref{fig3}) that such 
an approach does not either include the correct physics, or predict 
the magnitude of the effect.
The convolution approach served us however as a guidance.   
One can draw similar conclusions, in fact, by considering the case $P^2 \neq M^2$, 
where one an extra term dependent on nucleon off-shellness can be added to the expansion 
in Eq.(\ref{Taylor1}). 
This term has the same structure with respect to its $t$-dependence as the other
terms in the ``on-shell'' series. 
In other words, nucleon off-shellness further depletes the bound 
nucleon's momentum.   

An important conclusion of our study is, therefore, that details of the long range
nuclear interactions seem to have little influence on the bound nucleon GPDs.

\subsubsection{$t$-dependence as a Constraint on GPD Parametrizations}       
\begin{figure}
\vspace{3cm}
\includegraphics[width=8.cm]{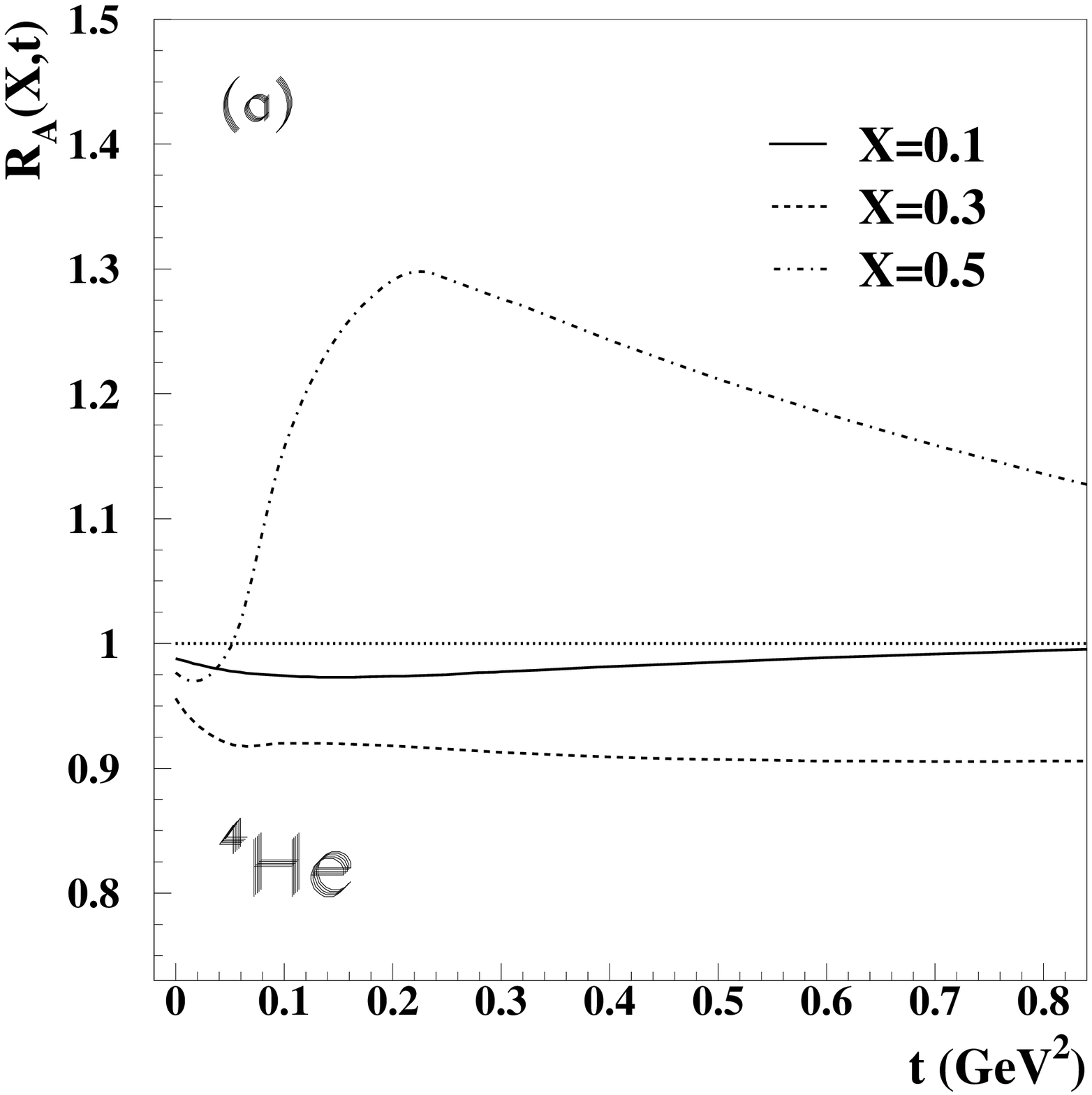}
\includegraphics[width=8.cm]{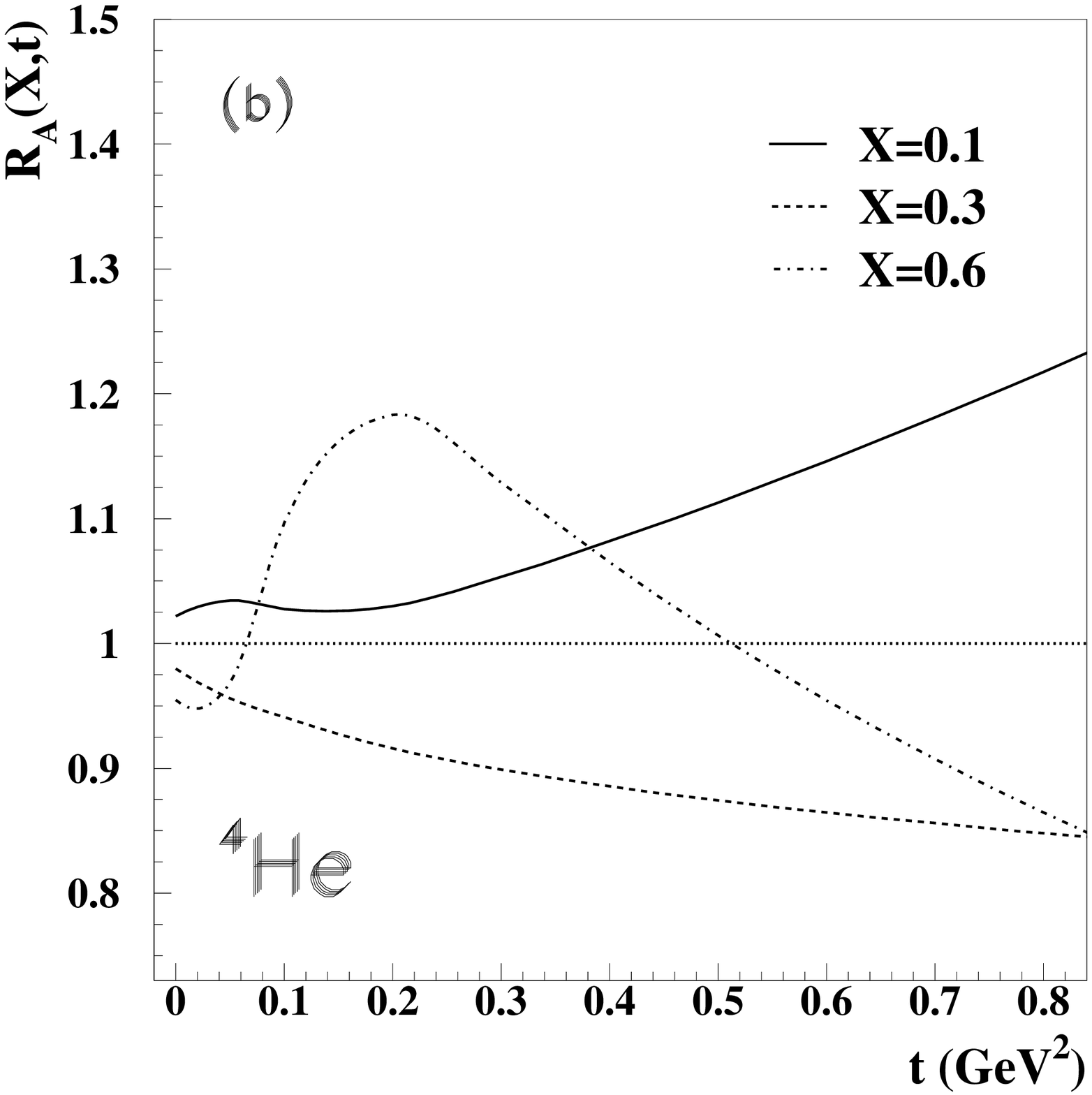}
\caption{$t$-dependence of the generalized EMC effect for different values of $X$:
$X=0.1$ (full), $X=0.3$ (short dashes), $X=0.6$ (dot-dashes). 
{\bf (a)} No off-shell effects; {\bf (b)} with off-shell effects. Notice the striking
difference between the off-shell and on-shell curves discussed in the text at both low 
and large values of $X$.} 
\label{fig8}
\end{figure}

In Fig.\ref{fig8} we show the $t$ dependence of the ratio $R^A$, 
for both the on-shell (left panel) 
and off-shell (right panel) cases, for three different values of $X$, $X=0.1,0.3,0.6$. 
One can observe markedly different behaviors in the slopes in the two panels. 
At $X =0.6$ and $t>0.1$ GeV$^2$ one can observe a decreasing slope both in the on-shell and off-shell
cases. This characteristic behavior provides a method to use nuclear effects to 
distinguish between
different parametrizations of GPDs.
Recently, a number of new models were proposed that improve on factorized 
ansatze of the type 
$H(X,\zeta,t) = H(X,\zeta) F(t)$, used in many earlier theoretical 
evaluations.   
The models are obtained for  $\xi (\zeta=0)$. include parametrizations based 
on an exponential fall-off 
with $t$ of the type \cite{Bur,Diehl_ff,Vand}:   
\begin{equation}
H(x,t) = f(x) \, \exp \left[ a(x) t \right]
\label{exp}
\end{equation}
By inspecting Eq.(\ref{RA}), one can conclude that:

\noindent  
{\it i)} No $t$ dependence of $R^A$ should be observed in a factorized model; 

\noindent
{\it ii)} The exponential models predict a linear dependence in $t$ originating from 
the $X$ derivative of $H^N$; 

\begin{figure}
\includegraphics[width=16.cm]{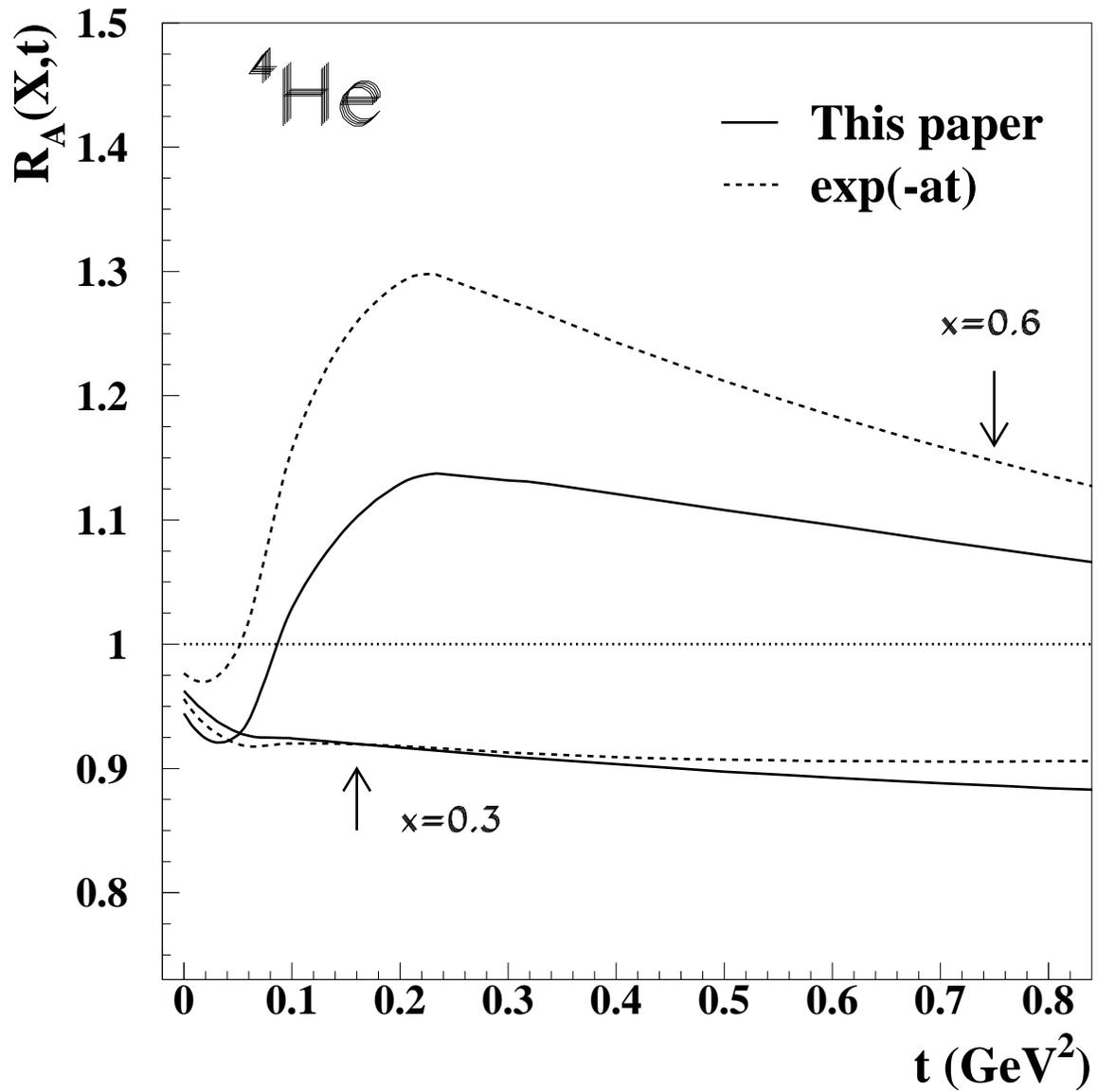}
\caption{$t-$-dependence of the generalized EMC effect for two different
types of parametrizations: The LC wave function one presented in this paper (full curves),
and the ``exponential'' type adopted by several groups (\cite{Bur,Vand,Diehl_ff} and 
references therein).} 
\label{fig9}
\end{figure}

\noindent 
{\it iii)} Other models predict a $t$ dependence that deviates 
from the linear behavior.
The comparison of exponential type parametrizations and the quark-diquark model
used in this paper is presented in Fig.\ref{fig9}.

\section{Sum Rules in Nuclei}
\label{moments-sec}
Mellin moments of order $n$ for the nucleon GPDs can be defined, analogously to the forward 
case, by integrating over $x$ the matrix elements in Eqs.(\ref{sfn}) and (\ref{convo1}), multiplied 
by $x^{n}$. 
\footnote{We use the notation in Eq.(\ref{kin1}) that is more appropriate
for the discussion of Mellin moments.}
Differenty from the forward case, the moments will depend on the variables $t$ and $\xi$, besides
displaying the expected $Q^2$ dependence. In other words, they can be written as $t$-dependent 
form factors of the 
same matrix elements of twist two local operators that appear in the forward case,
($t=0$). 
Special cases are given by the form factors ($n=0$):
\begin{eqnarray}
\int_{-1}^1 dx H_q(x,\xi,t) & = & F_1^q(t) \\ 
\int_{-1}^1 dx E_q(x,\xi,t) & = & F_2^q(t),
\label{1moment} 
\end{eqnarray}
where $q=u,d,s, ...$, and by the second moments ($n=1$), expressed through the 
symmetric energy-momentum tensor, 
in terms of the form factors of the quarks momentum, $M_2^q(t)$, and angular momentum, 
$J^q(t)$, respectively \cite{Ji1}:
\begin{eqnarray}
\int_{-1}^1 dx x H_q(x,\xi,t) & = & M_2^q(t) + \frac{4}{5} d_1(t) \xi^2 \\
\int_{-1}^1 dx x E_q(x,\xi,t) & = & (2 J^q(t) - M_2^q(t) ) - \frac{4}{5} d_1(t) \xi^2,
\label{2moment}
\end{eqnarray}
$M_2^q(0)$ is the total momentum fraction
carried by the quarks, $J^q(0)$ is the total angular momentum.
$d_1(t)$ is the moment of the first gegenbauer coefficient in the expansion of 
the $D$-term (see \cite{Die_rev} and references therein). Also, we do not write explicitly
the dependence on $Q^2$, unless necessary.  
\footnote{The original derivation of Eq.(\ref{2moment}) was in fact what motivated
the introduction of GPDs in Refs.\cite{Ji1,Rad1}.} 
Similar expressions exist for the gluon components. 

The sum rules in a nucleus can be constructed from Eqs.(\ref{1moment}) and (\ref{2moment}) as:
\begin{eqnarray}
\int_{-A}^A dx H^A(x,\xi,t) & = & F^A(t) \\ 
\int_{-A}^A dx x H^A(x,\xi,t) & = & M_2^A(t) + \frac{4}{5} d_1^A(t) \xi^2, 
\end{eqnarray}
In \cite{Poly} the connection was shown between 
$d_1^A(0)$ and the components of the energy-momentum tensor describing the
shear forces in the hadronic system. 
We can evaluate the Mellin moments
in a nucleus according to the microscopic approach described in the previous Sections. By restricting 
to the valence quarks contributions only ($x,y>0$), and by disregarding off-shell effects, 
one has the intuitive factorized form:
\begin{equation}
M_n^A(t) = \left( \int_0^A dy y^{n-1} f_A(y,t) \right) 
\left( \int_0^1 dx x^{n-2} [x H^N(x,\xi,t)] \right), 
\end{equation}
where $f_A(y,\xi,t) \approx f_A(y,\xi=0,t) \equiv f_A(y,t)$ in a nucleus, within
a non-relativistic description.
For $n=1$ and $n=2$ one has specifically:
\begin{eqnarray}
F^A(t) & = & F^{A, point}(t) F^N(t) \\
M_2^A(\xi,t) & = & M_2^{A, point}(t)M_2^N(t) + M_0^{A, point}(t)\frac{4}{5} d_1^N(t) \xi^2, 
\end{eqnarray}
with $M_n^{A, point}(t) = \int dy y^{n-1} f_A(y,t)$, the nuclear moment obtained 
by considering ``point-like'' nucleons. At $\xi=0$ one has:
\begin{equation}
M_2^A(t)  =  M_2^{A, point}(t) M_2^N(t), 
\end{equation}
related to the average value of the longitudinal momentum 
carried by the quarks
in a nucleus: 
\begin{eqnarray}
\langle x(t) \rangle_A  =  \frac{M_2^{A}(t)}{F^A(t)} 
 =  \frac{M_2^{A, point}(t)}{F^{A, point}(t)} \, \frac{M_2^{N}(t)}{F^N(t)}
 =  \langle y(t) \rangle_A \langle x(t) \rangle_N, 
\label{mom_convo}
\end{eqnarray}
The D-term in a nucleus reads: 
\begin{equation}
d_1^A(t)  =  M_0^{A, point}(t) d_1^N(t)
\label{Dterm_A}. 
\end{equation}
At $t=0$ one has: 
$d_1^A(0) \approx 1/[1-\overline{E}/M + 2/3 \langle P^2 \rangle /2M^2]_A d_1^N(0)$.
The microscopic calculation presented here predicts therefore an $A$-dependence of $d_1^A(0) \propto A$, 
modulo a $\propto \ln A$ enhancement from nuclear dynamics, {\it i.e.} from the 
$A$-dependent term multiplying $d_1^N(0)$. 
This is at variance with the estimate,
$d_1^A(0) \propto A^{7/3}$ obtained
by calculating directly the nuclear energy-momentum tensor in a liquid drop picture of the 
nucleus \cite{Poly}.      
Whether the two pictures can be reconciliated, that is addressing the question of whether the
shear forces are either under- or overestimated in either model, is an interesting problem for 
future investigations.

\begin{figure}
\includegraphics[width=16.cm]{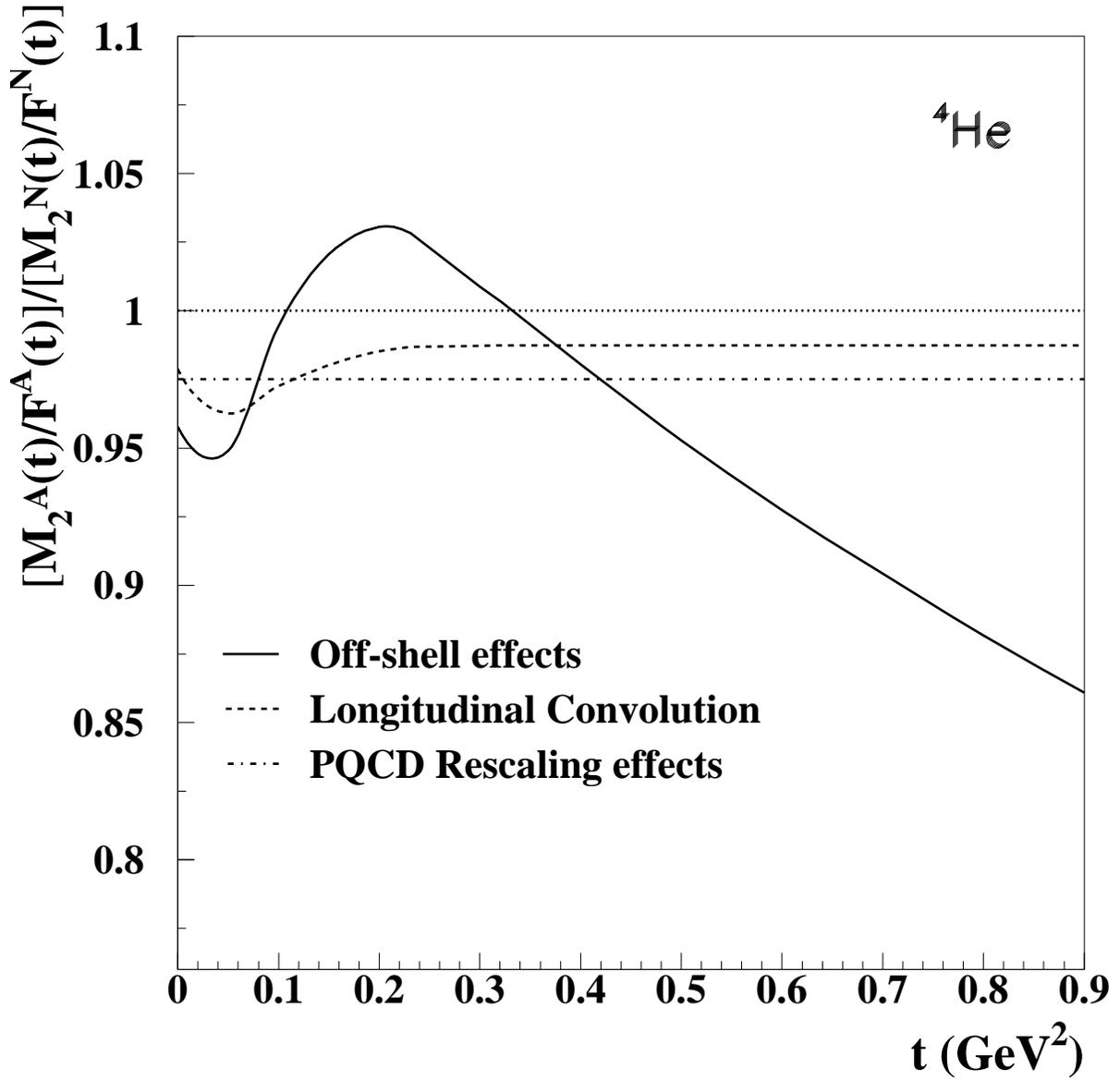}
\caption{Ratio of $n=2$ GPD moments of bound nucleons in $^4He$ over the free nucleon ones
as a function of $t$. 
Moments have been normalized to the form factors (see text). The intercept at $t=0$ represents
the ratio of the momentum fraction taken by valence quarks in a nucleus with respect to
the free nucleon one.} 
\label{fig10}
\end{figure}

Finally, we discuss how the moments of GPDs in nuclei might provide new insight on the origin of 
nuclear medium modifications of hadrons. Two distinct pictures based on hadronic and 
partonic degrees of freedom respectively have been put forth to explain the effect. 
On one side, the convolution formalism discussed here makes use of ``hadronic'' degrees of
freedom, although the hadronic structure can be modified due to off-shell effects. The nature 
of such modifications resides in the parameters of nucleon dynamics and interactions.     
On the other hand, other descriptions (\cite{Pirner} and references therein), relate the 
modifications of ``partonic'' parameters such as the string tension or the confinement
radius to density dependent effects in the nuclear medium. The authors of Ref.\cite{CRR} (CRR) 
in fact, even formulated the hypothesis of a ``duality'' scenario between the two pictures.      
A qualitatively different description that would help disentangle the two, can be obtained from 
the investigation of form factor (or spatial) type quantities
such as the ones described in Eqs.(\ref{1moment}-\ref{Dterm_A}).
The effect of partial deconfinement predicts a similar formula for $M_2^{A}(t,Q^2)$, $\xi=0$
calculated in the convolution formalism, namely:
\begin{eqnarray}
M_2^{A, CRR}(t,Q^2) & = & \left[ (1- d_{NS}^{n=2} \kappa_A) F^{A, point}(t) \right] M_2^N(t,Q^2),
\label{mom_CRR}
\end{eqnarray} 
$d_{NS}^{n=2}$ being the anomalous dimensions for the Non-Singlet ($NS$) 
sector, and $\kappa_A$ being the parameter
modifying the $Q^2$ scale dependence in Ref.\cite{CRR}. 
In Fig.\ref{fig10} we show the ratio of $M_2^{A}/F^{A}$ over $M_2^{N}/F^{N}$ in several
scenarios: the longitudinal convolution formula where the ratio coincides with $\langle y(t) \rangle_A$
from Eq.(\ref{mom_convo}); the ratio calculated including off-shell effects (that break the factorization
in Eq.(\ref{mom_convo})); and the prediction from Eq.(\ref{mom_CRR}). 
The most important aspect of this graph is that while ``factorization based'' approaches such
as the longitudinal convolution formalism and the CRR-model, predict a $t$ dependence of the
ratio that is approximately flat, off-shell effects have an impact on the $t$ dependence, and
are therefore indispensable for treating the transverse spatial dependence (or $3D$ imaging) 
in a nucleus, a subject to be explored in future studies \cite{LiuTan3}.     

Besides the results on the $NS$ contribution, yet another interesting possibility 
emerges from the non-forward scattering extension of nuclear deep inelastic 
structure studies.
Predictions were given in Ref.\cite{CRR} on the behavior of
the binding quanta appearing in the expression for the second moment for the sea quarks. 
A relation was found between the sea quarks, $q_s$ component of the 
``binding quanta'', $B$, $M_2^{q_s/B}$, and the gluon component of the nucleon $M_2^{G/N}$: 
\begin{equation}
M_2^{q_s/B}(Q^2)  \approx -\frac{3}{16}f  M_2^{G/N}(Q^2),
\end{equation}
from which it was deduced that the binding quanta, identified in \cite{CRR} 
with nuclear pions, are gluonic in nature, or in other words, they contain gluons 
that cannot be generated by evolution. GPDs provide a new method
to study the form factor of such gluon dominated particles, that could be identified 
as candidates for glueballs.
One can in fact use the perturbative QCD (PQCD) equations of Ref.\cite{CRR} 
in conjunction with the formalism described above to obtain:
\begin{equation}
M_2^{q_s/B}(t) = \frac{1}{M_2^{B/A}(t)} \left( M_2^{q_s/A}(t) - M_2^{q_s/N}(t) M_2^{N/A}(t) \right).
\end{equation}

\section{Conclusions}
\label{conc_sec}
We conducted an exploratory study using on one side GPDs as tools 
to unravel the deep inelastic transverse structure of nuclei, and on the
other obtaining information on GPDs themselves by observing how they 
become modified in the nuclear environment. 
We used a microscopic approach restricted mainly to the valence structure. 
We conclude that: 

\vspace{0.3cm}
\noindent{\it i)} Although GPDs probe transverse degrees of freedom
in a nucleus, through the variable $t$, modifications of the transverse momentum 
dependence are described similarly to the forward case, through the $k_\perp, P_\perp$ 
variables. These variables are however not Fourier conjugates to $\Delta$, thus rendering
the description of off-shell effects more subtle. 

\noindent{\it ii)} The unraveled $t$ dependence of nuclear modifications 
originates from different mechanisms at the
nuclear and nucleon vertices, respectively. By taking nuclear ratios of GPDs normalized 
to form factors, we demonstrated
that the details of the nuclear long range interactions (two-body currents, large distance
behavior of nuclear density, etc. ...) can be disregarded compared to the deep inelastic 
induced modifications of the bound GPDs. 
This result bears important consequences since it allows us on one side to avoid the 
intricacies and details related to the evaluation of nuclear form factors, and on the other,
it points at interesting physical phenomena determined by the short range part of the 
nuclear interactions.

\noindent{\it iii)} The pattern of nuclear modifications predicted, and in 
particular the deviations of off-shell effects from the longitudinal convolution
provide clear signals to be sought in experimental measurements. 

\noindent{\it iv)} Data on DVCS in nuclei will help one distinguish among models for GPDs. 
Distinctive behaviors in {\it e.g.} the $t$ dependence emerge for the factorized, the exponential based, 
and other LC based models presented here. 

\noindent{\it v)} Interesting relationships were found by studying Mellin moments in nuclei. In particular,
we predicted the $A$-dependence for the $D$-term of GPDs within a microscopic approach, and the behavior
with $t$ of the total momentum carried by quarks in a nucleus. By studying Mellin moments we were able
to make a connection with widely used approaches that relate the 
modifications of ``partonic'' parameters such as the string tension or the confinement
radius to density dependent effects in the nuclear medium \cite{Pirner,CRR}. 
We consider this an important result of our paper 
at the light of nuclear hadronization studies which are vital for interpreting current and future data 
at RHIC, HERMES and Jefferson Lab (\cite{Pirner} and references therein). 

Many questions remain, some of which will be addressed in a forthcoming paper \cite{LiuTan3}.
These will include a detailed study of the impact parameter ($b$) dependence, and 
predictions of quantities measurable at currently available energies, namely both beam and target 
asymmetries, including a quantitative evaluation of the coherent vs. incoherent nuclear contributions.

\acknowledgments
We thank Delia Hasch for discussing HERMES experiments, and Wally Melnitchouk for 
discussions.
This work is supported by the U.S. Department
of Energy grant no. DE-FG02-01ER41200. 

\appendix*
\section{Off-forward propagator}
We evaluate the trace in Eq.(\ref{matrix2}), that gives origin to the
square root factors in our model for nucleon GPDs, Eq.(\ref{HN}).

We start from defining the sum over the spins in Eq.(\ref{matrix2}), for particles
with different momenta, $k\equiv(E;{\bf k})$, and $k^\prime = k + \Delta$. The spinors are 
defined as: 
\begin{eqnarray}
u(k,s) = N (\not\!k + m) u(0,s),
\end{eqnarray}
with: $u_1(0,s) = 1$, $u_i(0,s)=0, i=2,3,4$, ($u_2(0,-s) = 1$, $u_i(0,-s)=0, i=1,3,4$);
and $\displaystyle N^{-1}=(E+m)^{1/2}$.   
In the forward case, therefore:
\begin{eqnarray}
\sum_s u(k,s)\bar{u}(k,s) = (\not\!k + m) 
\end{eqnarray}
In the off-forward case:
\begin{eqnarray}
\sum_s u(k,s)\bar{u}(k^\prime,s) & = & 
N N^\prime
\sum_s (\not\!k + m) u(0,s) \bar{u}(0,s)(\not\!k^\prime + m)   \nonumber \\
& =  & N N^\prime (\not\!k + m)\frac{(1+\gamma_o)}{2} (\not\!k^\prime + m) 
\end{eqnarray}
The trace of this quantity is given by:
\begin{eqnarray}
{\rm Tr} \left\{ \sum_s u(k,s)\bar{u}(k^\prime,s) \right\} & = & 2 N N^\prime
\left[(k.k^\prime + m^2) + m (k_0 + k_{0}^{\prime})\right]
\end{eqnarray}
In light cone coordinates the above is expressed as:
\begin{eqnarray}
{\rm Tr} \left\{ \sum_s u(k,s)\bar{u}(k^\prime,s) \right\} & = &
\frac{2}{\sqrt{(\frac{k^+ + k^-}{\sqrt{2}}+m)(\frac{k^{\prime +} + 
k^{\prime -}}{\sqrt{2}}+m)}} 
\left[ (k^+ k^{\prime -} + k^- k^{\prime +} - k_{\perp} k_{\perp}^{\prime}+ m^2) \right. \nonumber \\ 
& & \left. + \frac{m}{\sqrt{2}} ((k^+ + k^-) + ( k^{\prime +} + k^{\prime -}))\right]
\end{eqnarray}
thus, at leading order:
\begin{eqnarray}
{\rm Tr} \left\{ \sum_s u(k,s)\bar{u}(k^\prime,s) \right\}  = 
2 m \frac{(k^+ + k^{\prime +})}{\sqrt{k^+ k^{\prime +}}} =  
2 m \frac{2X - \zeta}{\sqrt{X(X-\zeta)}}
\label{trace0}
\end{eqnarray} 
and the trace with $\gamma^+$ appearing in Eq.(\ref{HN}), is given by:
\begin{eqnarray}
\frac{1}{2P^+} {\rm Tr} \left\{ \gamma^+ \sum_s u(k,s)\bar{u}(k^\prime,s) \right\}  = 
\frac{2}{P^+} \sqrt{k^+ k^{\prime +}} = 2 \sqrt{X} \sqrt{X-\zeta} 
\label{trace1}
\end{eqnarray}

\vspace{5cm} 
\newpage

\vspace{1cm}

\newpage

\newpage

\newpage

\newpage
\newpage

\end{document}